\documentclass[12pt]{iopart}

\expandafter\let\csname equation*\endcsname\relax
\expandafter\let\csname endequation*\endcsname\relax
\usepackage{amsmath}
\usepackage{amsfonts}
\usepackage{amssymb}
\usepackage{mathrsfs}
\usepackage{graphicx, epstopdf}
\usepackage{multirow}
\usepackage{color}

\begin{document}

\def\Q{\mathbf{Q}}
\def\Qvec{\mathbf{Q}}
\def\P{\mathbf{P}}
\def\I{\mathbf{I}}
\def\k{\mathbf{k}}
\def\z{\mathbf{z}}
\def\zhat{\mathbf{z}}
\def\n{\mathbf{n}}
\def\m{\mathbf{m}}
\def\r{\mathbf{r}}
\def\x{\mathbf{x}}
\def\J{\mathbf{J}}
\def\Rr{\mathbb{R}}
\def\u{\mathbf{u}}
\def\nvec{\mathbf{n}}
\def\V{\mathbf{V}}
\def\_v{\mathbf{v}}
\def\F{\mathbf{F}}

\title[Hierarchies of Critical Points of a Landau-de Gennes Free Energy]{Hierarchies of Critical Points of a Landau-de Gennes Free Energy on Three-Dimensional Cuboids}

\author{Baoming Shi$^1$, Yucen Han$^2$, Jianyuan Yin$^3$, Apala Majumdar$^2$ and Lei Zhang$^{4,*}$}

\address{$^1$ School of Mathematical Sciences, Peking University, Beijing 100871, China.}
\address{$^2$ Department of Mathematics and Statistics, University of Strathclyde, Glasgow, G1 1XH, UK.}
\address{$^3$ Department of Mathematics, National University of Singapore, Singapore 119076.}
\address{$^4$ Beijing International Center for Mathematical Research, Center for Quantitative Biology, Peking University, Beijing 100871, China.}

\ead{\mailto{ming123@stu.pku.edu.cn}, \mailto{yucen.han@strath.ac.uk}, \mailto{yinjy@nus.edu.sg}, \mailto{apala.majumdar@strath.ac.uk}, \mailto{zhangl@math.pku.edu.cn}}

\begin{abstract}
We investigate critical points of a Landau-de Gennes (LdG) free energy in three-dimensional (3D) cuboids, that model nematic equilibria. We develop a hybrid saddle dynamics-based algorithm to efficiently compute solution landscapes of these 3D systems. Our main results concern (a) the construction of 3D LdG critical points from a database of 2D LdG critical points and (b) studies of the effects of cross-section size and cuboid height on solution landscapes. In doing so, we discover multiple-layer 3D LdG critical points constructed by stacking 2D critical points on top of each other, novel pathways between distinct energy minima mediated by 3D LdG critical points and novel metastable escaped solutions, all of which can be tuned for tailor-made static and dynamic properties of confined nematic liquid crystal systems in 3D.
\end{abstract}


\section{Introduction}
Liquid crystals are mesophases intermediate between the solid and liquid states. The simplest liquid crystal phase is the nematic phase for which the constituent molecules have no positional order, but tend to align along certain locally preferred directions \cite{de1993physics}, referred to as \emph{nematic directors}. Consequently nematic liquid crystals (NLCs) have direction-dependent physical, optical and rheological properties \cite{sonin2018pierre,stewart2019static}. Thus, NLCs have widespread applications in opto-electronics, nanodevices and materials technologies. A crucial feature of NLC systems concerns topological (point or line) defects, interpreted as discontinuities in the directors, which have been visualized in polymeric materials or through mesoscale simulations of the local orientation of the molecules \cite{duclos2020topological}. The defects play important roles in self-assembled structures, colloidal suspensions and multistable systems, and often label families of observable equilibria and transient states in static and dynamic phenomena for confined NLC systems \cite{foffano2014dynamics,miller2014design,LAGERWALL20121387}.

There are multiple microscopic, mean-field and macroscopic/continuum theories for NLCs, e.g. Maier-Saupe, Oseen-Frank, Ericksen theories and the Landau-de Gennes (LdG) theory, which is the most powerful and general continuum theory amongst its competitors \cite{doi1988theory,wang2021modeling}. 
The LdG theory is a variational theory and describes the NLC state by a macroscopic order parameter, the $\Qvec$-tensor order parameter with five degrees of freedom, and the physically observable NLC states as minimizers of an appropriately defined LdG free energy, subject to physically relevant boundary conditions. Of course, the LdG free energy typically has multiple energy minimizers and non energy-minimizing critical points, all of which make the mathematics and physics of NLCs challenging and fascinating. The precise details are given in the next section, but there has been substantial recent work on the reduced LdG model, valid for two-dimensional (2D) confinement and for planar director profiles \cite{canevari2017order,robinson2017molecular,wang2019order,han2021elastic}. In this reduced LdG framework, there are only two degrees of freedom and the reduced LdG energy effectively reduces to the celebrated Ginzburg-Landau energy for superconductors \cite{han2020reduced}. There has been a body of work for reduced LdG critical points on a 2D square domain with tangent boundary conditions, motivated by the experimental work in \cite{tsakonas2007multistable}. For small squares on the nano-scale, there is a unique reduced LdG critical point, coined as the Well Order Reconstruction Solution (WORS), which has a pair of orthogonal line defects along square diagonals \cite{kralj2014order}. The WORS is globally stable when the square edge length is sufficiently small, but loses stability as the edge length increases \cite{canevari2017order}. For a large square domain, there are two types of experimentally observed stable states: the diagonal (D) state for which the director is aligned along the square diagonal and the rotated (R) state for which the director rotates by $\pi$ radians between a pair of opposite square edges \cite{tsakonas2007multistable}. In \cite{yin2020construction}, the authors investigate the solution landscape of the reduced LdG model on square domains, and recover the typical WORS, D, and R states, along with new unstable states that have multiple point/line defects, and the switching mechanisms between them. More generally, reduced LdG models have been studied on 2D polygons such as a hexagon, 2D discs and rectangles, and the reader is referred to \cite{hu2016disclination,han2020solution, fang2020surface,shi2021nematic}.

The 2D studies cited above are limiting cases  of 3D studies, with vanishing height \cite{han2020reduced}. This raises the fundamentally important question - reduced 2D studies only exploit two out of five degrees of freedom in the LdG framework and how do the additional degrees of freedom manifest in 3D? From an application point of view, 3D studies are much needed in generic scenarios such as liquid crystal displays, food science, and biology \cite{doi:10.1126/science.283.5398.57}.  In general, 2D solutions (or critical points of a reduced LdG free energy) can be viewed as $z$-invariant 3D LdG critical points, invariant in the third dimension. With an additional dimension in 3D, we have the possibility of 3D $z$-variant solutions with complicated defect structures \cite{duclos2020topological,long2021geometry}, more complicated solution landscapes with $z$-variant 3D stable and unstable critical points, $z$-variant pathways between different critical points and far greater tunability of solution landscapes for designer material properties. 
In \cite{canevari_majumdar_wang_harris}, the authors report a mixed solution in a 3D cuboid that interpolates between two distinct stable D states, on the top and bottom cuboid surfaces. In a cylinder, we have a 3D escaped solution with two ring disclinations, and the domino-like transition pathway mediated by a $z$-variant unstable LdG critical point is energetically preferable to the $z$-invariant pathways \cite{han2019transition}. Various 3D knotted defect fields in confined NLCs are shown in \cite{Knotteddefects}, which cannot be captured by 2D studies alone. The authors of \cite{duclos2020topological} report the experimental visualization of the defect structure, and demonstrate the continuous switching between a $+1/2$ point defect and a $-1/2$ defect by twisting along $z$-direction, again outside the remit of 2D studies. These genuinely 3D features of confined NLCs motivate us to systematically study LdG solution landscapes, with the full five degrees of freedom, on a 3D cuboid as a generic example, by using 2D critical points in \cite{robinson2017molecular,yin2020construction} as a solution database. 

More precisely, we focus on critical points of a LdG free energy on a 3D cuboid, with the full five degrees of freedom, that model nematic equilibria and admissible nematic states, imposing tangent Dirichlet boundary conditions on lateral surfaces and natural boundary condition on top and bottom surfaces. There are two geometry-dependent variables: the edge length of the square cross-section denoted by $\lambda$, and the cuboid height denoted by $h$. Our goal is to use the database of 2D LdG critical points in \cite{robinson2017molecular,yin2020construction} (for a square domain) to systematically construct both $z$-invariant and $z$-variant critical points of a 3D LdG energy. 
In doing so, we find that many $z$-variant solutions have inherently small eigenvalues for the Hessian of the LdG energy, reflected in the insignificant energy cost of moving cross-sectional solution profiles up and down, provided $h$ is large enough. We design a hybrid numerical scheme to deal with the ill conditioned saddle dynamics and convergence issues, caused by such small eigenvalues. This hybrid numerical scheme for the saddle dynamics allows us to efficiently explore the solution landscapes of this 3D system as a function of $\lambda$ and $h$, with special attention to the elusive unstable LdG critical points. 
Our first numerical result concerns the 3D $z$-invariant critical points that are a translationally invariant version of the 2D reduced LdG critical points. These 2D critical points survive as $z$-invariant solutions in 3D but are \emph{more unstable} in 3D, i.e. they have higher Morse indices or equivalently, more unstable directions in 3D compared to 2D. 
Our main results concern new 3D LdG critical points, labelled as A1-B-A2, where the labels A1, B and A2 come from the 2D LdG critical points (critical points of the reduced LdG energy), which are approximated by the profiles on the top, middle, and bottom slices of the $z$-variant 3D critical point. We can use the pathways, A1 $\rightarrow$ B $\rightarrow$ A2 on the 2D solution landscape, where the reduced 2D LdG critical point B usually has a higher Morse index than A1 and A2, to construct candidates for 3D LdG critical points, labelled by A1-B-A2.
We also observe the emergence of multiple-layer solutions, which are effectively blocks of dual 3D LdG critical points (A1-B-A2 and A2-B-A1) stacked on top of each other, and the Morse indices of these multiple-layer solutions depend on the number of layers.
We believe these numerical results to be of wide interest, since they provide a general recipe (which could fail in some situations) for constructing higher-dimensional critical points of a free energy from lower-dimensional critical points.
The recipe is intuitive but the plethora of numerical results, the symmetries of the 3D LdG critical points and their defect sets give great inroads into cutting-edge computational and modelling questions. 
There are some interesting by-products of these numerical experiments, which could be relevant for novel NLC applications engineered with 3D cuboids. We explore 3D nematic solution landscapes as outlined above and in doing so, find an energetically favourable pathway between two $z$-invariant energy-minimizing D states, and this pathway is featured by a $z$-variant transition state, for large enough $\lambda$ and $h$. Thus, $z$-variant critical points can be relevant for the switching between $z$-invariant states, which is interesting in its own right.

We numerically compute bifurcation diagrams for the 3D LdG critical points, as a function of $\lambda$ and $h$, which show that $z$-invariant solutions become more unstable while some $z$-variant solutions become more stable, as $h$ increases.  Whilst we solve for the full five degrees of freedom for the LdG $\Qvec$-tensor and allow for all variables to depend on all three spatial dimensions, the majority of our numerical results only have three degrees of freedom and the $z$-variant critical points emerge from the $z$-dependence of the degrees of freedom or the $z$-dependence of the nematic directors i.e. the nematic directors lie in the $xy$-plane but are not invariant in the $z$-direction. In the last sub-section, we numerically find a branch of escaped solutions for which the directors are out-of-plane, and which exploit the full five degrees of freedom 
and investigate the transition pathway between a stable escaped LdG critical point and the $z$-invariant D state.   

This paper is organized as follows. In Sec. 2, we briefly review the LdG theory for NLCs and introduce the domain and the boundary conditions. In Sec. 3, we propose a hybrid numerical scheme for the saddle dynamics to speed up the computation of saddle points. In Sec. 4, we present a detailed study of the 3D LdG model on cuboid. We finally present our conclusions in Sec. 5.

\section{The Landau--de Gennes theory}
We work within the celebrated LdG theory,  which is the most general continuum theory for nematic liquid crystals (NLCs). The LdG theory describes the NLC state by a macroscopic order parameter, the LdG $\Q$-tensor order parameter, that distinguishes NLCs from isotropic liquids in terms of anisotropic macroscopic quantities, such as the magnetic susceptibility and dielectric anisotropy \cite{de1993physics}. Mathematically, the $\Q$-tensor is given by a symmetric, traceless $3\times3$ matrix 
as shown below:
\begin{equation}\label{eq:5-degree}
    \Q = \begin{pmatrix}
        q_1-q_3 & q_2 & q_4\\
        q_2 & -q_1-q_3 &  q_5\\
        q_4 & q_5 &  2q_3
    \end{pmatrix}.
\end{equation}
From the spectral decomposition theorem, we can write the $\Q$-tensor as
$$
\Q = \sum_{i=1}^{3} \lambda_{i} \mathbf{e}_i \otimes \mathbf{e}_i,
$$
where $\left\{ \mathbf{e}_1, \mathbf{e}_2, \mathbf{e}_3 \right\}$ are the eigenvectors of the $\Q$-tensor and $\lambda_1\leqslant\lambda_2\leqslant\lambda_3$ are the associated eigenvalues respectively, subject to $\sum_{i=1}^{3}\lambda_i = 0$. The eigenvectors model the preferred directions of spatially averaged local molecular alignment in space or the nematic directors, and the eigenvalues are a measure of the degree of orientational order about these directions. A $\Q$-tensor is said to be isotropic if $\Q=\mathbf{0}$, uniaxial if $\Q$ has a pair of repeated non-zero eigenvalues, and biaxial if $\Q$ has three distinct eigenvalues \cite{de1993physics,mottram2014introduction}. Physically, a uniaxial NLC phase has a single distinguished direction of averaged molecular alignment, such that all directions perpendicular to the uniaxial director are physically equivalent. A biaxial phase has a primary and secondary nematic director. 

The LdG theory is a variational theory, based on the premise that the physically observable configurations are modelled by minimizers of an appropriately defined LdG free energy \cite{de1993physics}. There are several forms of the LdG free energy, and in this manuscript we work with a particularly simple form:
\begin{equation}
	E[\Q]: = \int_{V} \left[\frac{L}{2}\left| \nabla \Q \right|^2 + f_B\left( \Q \right)\right]\mathrm{d}A, 
\end{equation}
where the first term in the integrand is the Dirichlet elastic energy density that penalizes spatial inhomogeneities, and the second term is the thermotropic potential, $f_B$ that dictates the preferred NLC phase as a function of temperature. 
\begin{equation}\label{f_B}
	\left| \nabla \Q \right|^2:=\frac{\partial Q_{ij}}{\partial r_k}\frac{\partial Q_{ij}}{\partial r_k},\ i,j,k = 1,2,3,\text{  }
	f_B(\Q): = \frac{A}{2}\mathrm{tr} \Q^2 - \frac{B}{3} \mathrm{tr} \Q^3 + \frac{C}{4} (\mathrm{tr} \Q^2)^2-f_{B,0}.
\end{equation}
More precisely, the working domain is a cuboid $V=\left[ -\lambda,\lambda\right]^2 \times \left[-\lambda h,\lambda h \right]$ where $\lambda$ is edge-length of the 2D square cross-section and $h$ is a measure of the height ($h>0$); $L>0$ is a material-dependent elastic constant, $A=\alpha (T - T^*)$ is the rescaled temperature, with $\alpha>0$ and $T^*$ is a characteristic liquid crystal temperature; $B, C>0$ are material-dependent bulk constants. The minimizers of $f_B$ depend on $A$ and determine the NLC phase for spatially homogeneous samples. When $A> \frac{B^2}{24 C}$, the minimizer of $f_B$ is the isotropic state, and for $A<0$, the minimizers of $f_B$ constitute a continuum of uniaxial $\Q$-tensors defined by
\[
\mathcal{N} = \left\{ \Q = s_+\left(\mathbf{n}\otimes \mathbf{n} - \frac{\mathbf{I}}{3} \right) \right\},
\]
where
$$ s_+ = \frac{B + \sqrt{B^2 - 24 AC}}{4C}, $$
and
$\mathbf{n}$ is an arbitrary unit vector field that models the uniaxial director. The constant, $f_{B,0}=\frac{A}{3}s_+^2-\frac{2B}{27}s_+^3+\frac{C}{9}s_+^4$ \cite{majumdar2010equilibrium}, is added to ensure a non-negative energy density.

By rescaling the system according to $(\bar{x},\bar{y},\bar{z}) = (\frac{x}{\lambda},\frac{y}{\lambda},\frac{z}{\lambda})$, $\bar{\lambda}^2=\frac{2C\lambda^2}{L}$ and dropping the bars in subsequent discussions (so that all results are in terms of dimensionless variables), the non-dimensionalized LdG free energy is given by,
\begin{equation}
	E[\Q]:=\int_{V} \left[\dfrac{1}{2}\left|\nabla \Q \right|^2 +\lambda^2 \left(\frac{A}{4C}\text{tr}\Q^2-\frac{B}{6C}\text{tr}\Q^3+\frac{1}{8}(\text{tr}\Q^2)^2-\frac{f_{B,0}}{2C}
	\right) \right]\mathrm{d}A.
	\label{energy_b}
\end{equation}
The normalized domain is $V=\Omega\times \left[-h,h \right]$, $\Omega=\left[ -1,1\right]^2$ is the two-dimensional cross-section of the cuboid, and $\lambda^2$ describes the cross-sectional size. In what follows, we take fixed values of the parameters $B=0.64\times10^4 \text{Nm}^{-2}$,  $C=0.35\times10^4 \text{Nm}^{-2}$, and $L=4\times10^{-11} \text{N}$, which roughly correspond to the commonly used NLC material, MBBA \cite{wojtowicz1975introduction,majumdar2010equilibrium}. We focus on a special temperature $A=-B^2/3C$, which is a representative low temperature, to largely facilitate comparison with 2D results in \cite{robinson2017molecular,yin2020construction}. If the LdG critical point $\Q(x,y,z)$ depends on $z$, the 3D solution is $z$-variant. If $\Q(x,y,z)$ only depends on $x$ and $y$, i.e., $\Q(x,y)$, the 3D solution is $z$-invariant.

Of prime importance are nematic defects which have distinct optical signatures under a polarizing microscope \cite{de1993physics}. Motivated by the results in \cite{han2020reduced}, we use an innovative measure to identify defects. At the special temperature $A=-B^2/3C $, we have a branch of LdG critical points, $\Qvec_c$, with $q_4 = q_5 =0$ and constant $q_3 = -\frac{B}{6C}$, i.e.

\begin{equation}
    \label{Q_c}
\Q_c = q \left(\nvec_1 \otimes \nvec_1 - \nvec_2\otimes \nvec_2 \right) - \frac{B}{6C}\left(2 \zhat\otimes \zhat - \nvec_1 \otimes \nvec_1 - \nvec_2 \otimes \nvec_2 \right),
\end{equation}
where $\nvec_1$ and $\nvec_2$ are two orthogonal unit-vectors in the cuboid cross-section, and $\zhat$ is the unit-vector in the $z$-direction \cite{wang2019order}. These critical points only have $2$ degrees of freedom: $q$ and a degree of freedom associated with $\nvec_1$. The nematic director is defined to be the eigenvctor with the largest positive eigenvalue and the defect set is identified with the nodal set of $q$ i.e. a set of no order in the cross-sectional planes of the cuboid. Whilst solving for all five degrees of freedom, we numerically recover a class of critical points $\Q_c$ with only two degrees of freedom as above, and the $z$-dependence of $q(x,y,z)$ and $\nvec_1(x,y,z), \nvec_2(x,y,z)$ generate the novel $z$-variant 3D LdG critical points $\Q_c(x,y,z)$. When $q$, $\nvec_1$ and $\nvec_2$ do not depend on $z$, i.e., $q(x,y)$, $\nvec_1(x,y)$ and $\nvec_2(x,y)$, we recover the z-invariant 3D LdG critical points $\Q_c(x,y)$, constructed by reduced 2D LdG critical points reported in a batch of papers, on polygonal domains \cite{han2021elastic,han2020reduced}.
Hence, in these cases, 
we  use $\lambda_3-B/6C$ ($\lambda_3$ is the maximum eigenvalue of $\Q$) to visualize the location of defects \cite{yin2020construction,han2020solution,shi2021nematic} in Sections 4.1-4.5, and the zero set of $\lambda_3-B/6C$ labels the NLC defects.

For critical points with out-of-plane directors that exploit the full five degrees of freedom, defects can be tracked by the isosurface of biaxiality parameter $\beta^2=1-6\text{tr}(\Q^3)^2/(\text{tr}(\Q^2))^3$, $0\leqslant \beta \leqslant 1$. We have $\beta^2=0$ if and only if $\Q$ is uniaxial or isotropic \cite{landau2010beta}, and hence, we use the biaxiality parameter, $\beta^2$ to track defects in Sec. 4.6, which focuses on escaped critical points.

\begin{figure}[hbtp]
    \centering
    \includegraphics[width=.8\textwidth]{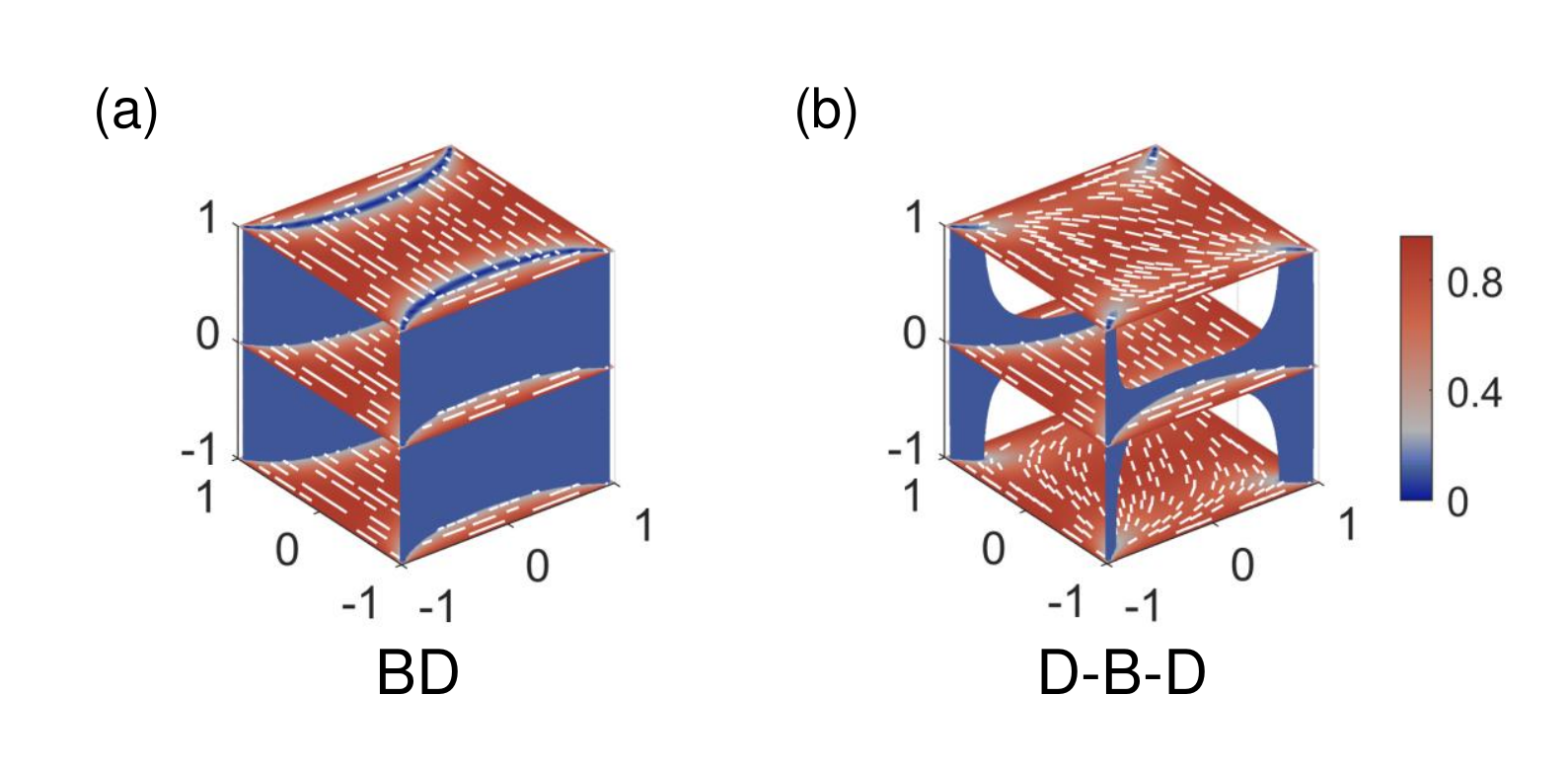}
	\caption{The plot of (a) $z$-invariant boundary distortion (BD) state and (b) $z$-variant D-B-D state \cite{canevari_majumdar_wang_harris}. The label D-B-D indicates that the state exhibits (approximate) D (Diagonal), BD (Boundary Distortion), and D profiles on the top, middle, and bottom slices, respectively.  The color bar and the white lines label the order parameter ($\lambda_3-B/6C$) and the director respectively. The blue regions ($\lambda_3-B/6C<0.1$) identify the NLC defects and are omitted in some following figures for better visualization.}
    \label{fig:1}
\end{figure}

With regards to boundary conditions, we use tangent Dirichlet boundary conditions on the lateral surfaces and Neumann boundary conditions on the top and bottom surfaces of the cuboid. The tangent Dirichlet conditions require the leading nematic director (with the largest positive eigenvalue) to be tangent to the edges of the 2D cross-section, $\Omega$ of the 3D cuboid.
This creates a natural mismatch at the four vertices of $\Omega$. Following the linear interpolation approach in \cite{han2020reduced,shi2021nematic,luo2012multistability}, we define the Dirichlet condition, $\Q=\Q_{bc}$ on the lateral surfaces, $x=\pm 1$ and $y=\pm 1$, in terms of a function with a shape parameter,
\begin{equation}
    \begin{cases}
    &\Q_{bc}(x=\pm 1,y,z) =\frac{s_+}{3}
    \left(
    \begin{tabular}{ccc}
        $-T_\epsilon(y)$  & $0$ &$0$ \\
        $0$  & $2T_\epsilon(y)$ &$0$\\
          $0$ &$0$  &$-T_\epsilon(y)$
    \end{tabular}
    \right),\\
    &\Q_{bc}(x,y=\pm 1,z) =\frac{s_+}{3}
    \left(
    \begin{tabular}{ccc}
        $2T_\epsilon(x)$  & $0$ &$0$ \\
        $0$  & $-T_\epsilon(x)$ &$0$\\
          $0$ &$0$  &$-T_\epsilon(x)$
    \end{tabular}
    \right),
    \end{cases}
    \label{Dirichlet bc}
\end{equation} 
where
\begin{equation}
    T_\epsilon(t)=\begin{cases}
        (1+t)/\epsilon, \ -1\leqslant t \leqslant -1+\epsilon,\\
        1, \ |t|\leqslant 1-\epsilon,\\
        (1-t)/\epsilon, \ 1-\epsilon\leqslant t \leqslant 1. 
    \end{cases}
\end{equation}
We take a sufficiently small $0<\epsilon\ll 1$, and the qualitative solution profiles are not changed by the choice of the interpolation.
The Neumann boundary conditions 
\begin{equation}\label{Natural bc}
    \frac{\partial\Q}{\partial n}= 0,\ z = \{-h,h\},\ (x,y)\in\Omega,
\end{equation}
where $n$ is the normal vector,
allow for $z$-invariant states i.e. NLC states which are invariant across the height of the cuboid (see  Fig. \ref{fig:1}(a)).

The critical points of the LdG free energy in (\ref{energy_b}) are classical solutions of the corresponding Euler-Lagrange equations:
\begin{equation}
\Delta \Q=\lambda^2\left(\frac{A}{2C}\Q-\frac{B}{2C}\left(\Q^2-\frac{tr(\Q^2)}{3}\I \right)+\frac{1}{2}tr(\Q^2)\Q\right),
\label{EL_b}
\end{equation}
with the imposed boundary conditions on the lateral surfaces \eqref{Dirichlet bc}, and natural boundary conditions in \eqref{Natural bc}.
The energy minimizers model the physically observable states, and there are a plethora of non energy-minimizing solutions of (\ref{EL_b}). 
In what follows, we study the relationships between the non energy-minimizing and energy-minimizing solutions of (\ref{EL_b}), and how the solution connectivity can be used to construct 3D NLC configurations on a cuboid. The cuboid is a generic and physically relevant example and our methods can be generalized to arbitrary 3D geometries.

\section{Numerical method}\label{numerical method}

In this section, we describe the numerical methods used to compute the critical points of the LdG free energy in (\ref{energy_b}), with special attention to the non energy-minimizing critical points which are typically hard to find. The critical points, $\Q$, are solutions of the Euler-Lagrange equations (\ref{EL_b}), which are a system of five nonlinear partial differential equations, for the five components of the $\Q$-tensor in \eqref{eq:5-degree}
and we solve for all $5$ degrees of freedom, $q_1, \ldots, q_5$.

A critical point of the LdG free energy, $\hat{\Q}$ is stable if the Hessian of the associated LdG free energy, $\nabla ^2 E(\hat{\Q})$, has only positive eigenvalues, and unstable if it has a negative eigenvalue. We study unstable saddle points of the LdG free energy, that are unstable in specific eigendirections. More precisely, for a non-degenerate index-$k$ (Morse index) saddle point $\hat{\Q}$, the Hessian $\nabla ^2 E(\hat{\Q})$ has exactly $k$ negative eigenvalues: $\lambda_1 \leqslant \cdots \leqslant \lambda_k$, corresponding to $k$ unit eigenvectors $\hat{\_v}_1,\cdots,\hat{\_v}_k$ subject to $\big\langle{\hat{\_v}_i}, \hat{{\_v}}_j \big\rangle = \delta_{ij}$, $1\leqslant i, j \leqslant k$. A stable critical point $\hat{\Q}$ is an index-0 critical point, i.e., the smallest eigenvalue of $\nabla^2 E(\hat{\Q})$ is positive. While a stable state can be relatively easily found by gradient descent method using a proper initial guess, finding a transition state (an index-$1$ saddle point) or high-index saddle points is much more difficult.
There are numerical methods for the computation of transition pathways mediated by index-$1$ saddle points, e.g. string methods \cite{weinan2002string,weinan2007simplified}, but they largely depend on a proper initial guess. However, initial guesses for saddle points are not easy to find since we typically do not have a priori knowledge of saddle points on the energy landscape. In what follows, we review the method of saddle dynamics and propose a hybrid numerical scheme to circumvent numerical stiffness and convergence issues.

\subsection{Saddle dynamics}
The saddle dynamics (SD) method \cite{2019High,yin2020constrained,zhang2021optimal} has been successfully used to efficiently compute the LdG critical points on 2D domain
\cite{han2020solution,shi2021nematic,yin2022solution}. We review the SD method in the following.
The SD for an index-$k$ saddle point $\Q$ (denoted by $k$-SD) is defined to be,
\begin{equation}
    \left\{
    \begin{aligned}
		\dot{\Q}&=- (\I-2\sum_{i=1}^k {\_v}_i{\_v}_i^\top)\nabla E(\Q), \\
		  \dot{\_v}_i&=-   (\I-{\_v}_i{\_v}_i^\top-\sum_{j=1}^{i-1}2{\_v}_j{\_v}_j^\top)\nabla^2 E(\Q) \_v_i,\ i=1,2,\cdots,k ,\\
    \end{aligned}
    \right.
	\label{eq: SD}
\end{equation}
where $\I$ is the identity operator. To avoid evaluating the Hessian of $E(\Q)$, we use the dimer
\begin{equation}
    h(\Q,\_v_i)=\frac{\nabla E(\Q+l\_v_i)-\nabla E(\Q-l\_v_i)}{2l}
\end{equation}
as an approximation of $\nabla ^2 E(\Q)\_v_i$, with a small dimer length $2l$. By setting the $k$-dimensional subspace $\mathcal{V}=\text{span} \big \{ \hat{\_v}_1,\cdots,\hat{\_v}_k \big \}$, $\hat{\Q}$ is a local maximum on $\hat{\Q}+\mathcal{V}$ and a local minimum on $\hat{\Q}+\mathcal{V}^\perp$, where $\mathcal{V}^\perp$ is the orthogonal complement of $\mathcal{V}$. The dynamics for $\Q$ in \eqref{eq: SD} can be written as
\begin{equation}
	\begin{aligned}
	\dot{\Q}&=\left(\I-\sum_{i=1}^k {\_v}_i{\_v}_i^\top\right) \left(-\nabla E(\Q)\right)+ \left(\sum_{i=1}^k {\_v}_i{\_v}_i^\top\right) \nabla E(\Q) \\
	&= \left( \I-\mathcal{P}_{\mathcal{V}}  \right)\left(-\nabla E(\Q)\right)+ \mathcal{P}_{\mathcal{V}} \left(\nabla E(\Q)\right),
\end{aligned}
\end{equation}
where $\mathcal{P}_{\mathcal{V}}\nabla E(\Q)=\left(\sum_{i=1}^k {\_v}_i{\_v}_i^\top\right)\nabla E(\Q)$ is the orthogonal projection of $\nabla E(\Q)$ on $\mathcal{V}$. Thus, $\left( \I-\mathcal{P}_{\mathcal{V}}  \right)\left(-\nabla E(\Q)\right)$ is a descent direction on $\mathcal{V}^\perp$, and $\mathcal{P}_{\mathcal{V}} \left(\nabla E(\Q)\right)$ is an ascent direction on $\mathcal{V}$. 

The dynamics for $\_v_i, i=1,2,\cdots,k$ in \eqref{eq: SD} can be obtained by minimizing the $k$ Rayleigh quotients simultaneously with the gradient type dynamics,
\begin{equation}
	\min_{{\_v}_i}\text{  }\left<\_v_i, \nabla ^2 E(\Q)\_v_i\right>,\ \text{s.t.}\ \left<\_v_i,\_v_j\right>=\delta_{ij}, \ j=1,2,\cdots,i,
\end{equation}
which generates the subspace $\mathcal{V}$ by computing the eigenvectors corresponding to the smallest $k$ eigenvalues of $\nabla^2 E (\Q)$.

\subsection{Hybrid numerical scheme}
\begin{figure}[hbtp]
    \centering
    \includegraphics[width=.6\textwidth]{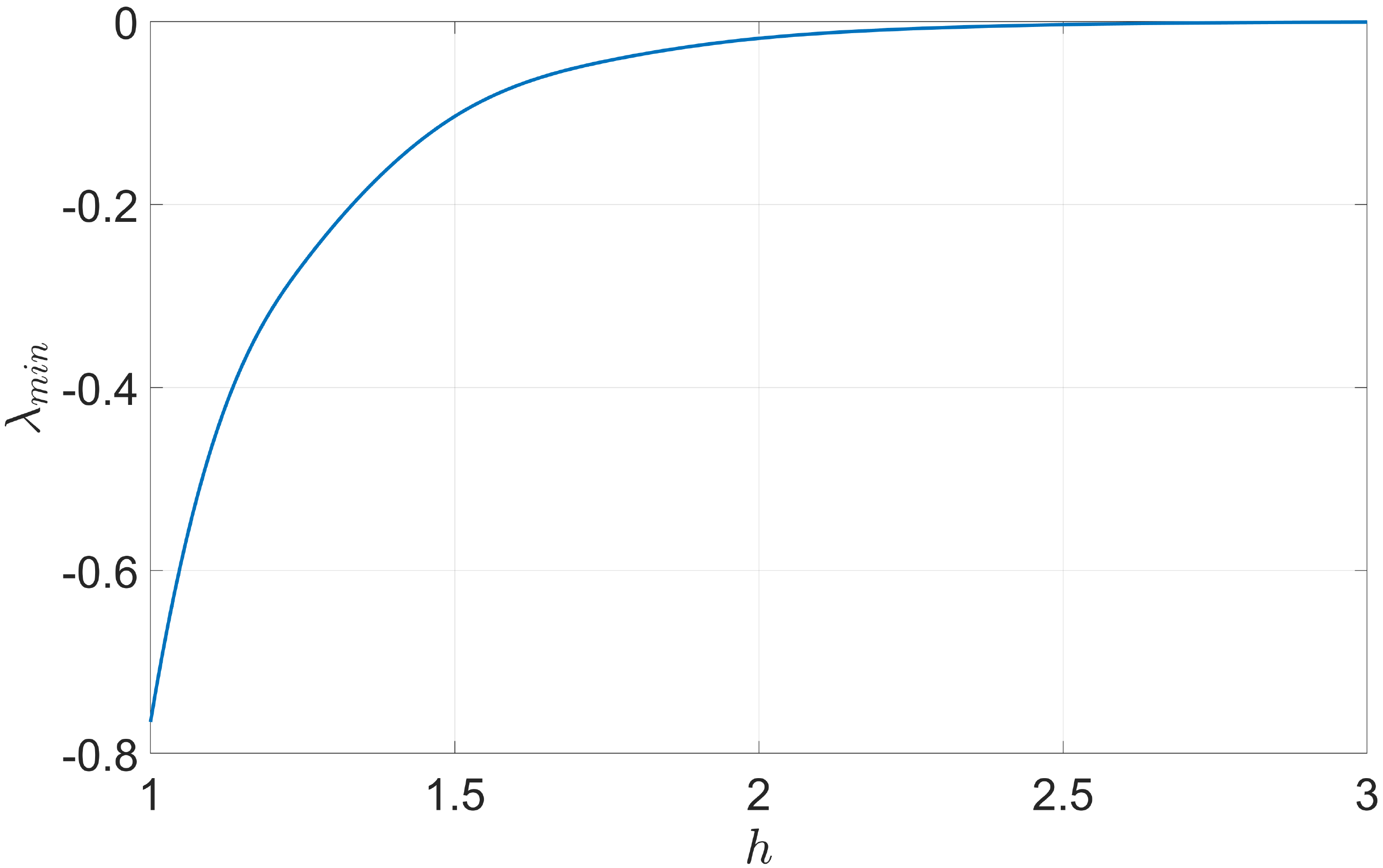}
    \caption{The smallest eigenvalue of D-B-D versus $h$ at $\lambda^2=30$.}
    \label{fig:2}
\end{figure}

We label hierarchies of LdG saddle points in a 3D cuboid by A-B-C, where A, B and C are the reduced 2D LdG critical points, approximated by the 2D profiles on $z=h$,  $z=0$ and $z=-h$ slices of a 3D LdG critical point. We find that $\nabla^2E(\text{A-B-C})$ usually have small eigenvalue with a large $h$, because we can move the middle state on $z=0$ up and down, without a significant energetic cost. If one eigenvalue, $\lambda_{\min}$ is close to zero (Fig. \ref{fig:2}), then this will cause numerical issues including the stiffness and slow convergence of the saddle dynamics. 

We elaborate on the numerical issues further by using $k$-saddle dynamics to find a target saddle point $\Q^*$, for which the smallest absolute eigenvalue, $\lambda_1$ is such that $|\lambda_1|<\epsilon$ and we consider the Jacobian operator of $k$-saddle dynamics,
\begin{equation}
    \J(\Q,\_v_1,\cdots,\_v_k)=\frac{\partial(\dot{\Q},\dot{\_v}_1,\cdots,\dot{\_v}_k)}{\partial(\Q,\_v_1,\cdots,\_v_k)}=\begin{pmatrix}
        \J_{\Q} & \J_{\Q 1} &\J_{\Q 2} & \cdots & \J_{\Q k}\\
        \J_{1\Q} & * &  \mathbf{0}& \cdots & \mathbf{0}\\
        \J_{2\Q} & * &  *& \cdots & \mathbf{0}\\
        \vdots &\vdots & \vdots& &\vdots \\
        \J_{k\Q}&* &* &\cdots &*
    \end{pmatrix}
\end{equation}
where 
\begin{equation}
    \J_{\Q}=\frac{\partial \dot{\Q}}{\partial \Q}=-\left (\I-\sum_{i=1}^{i=k}2\_v_i \_v_i^\top\right)\nabla^2 E(\Q),
\end{equation}
\begin{equation}
    \J_{\Q i}=\frac{\partial \dot{\Q}}{\partial \_v_i}=-2\left(\_v_i ^\top \nabla E(\Q) \I +\_v_i \nabla E(\Q)^\top \right).
\end{equation}
Now, we consider the spectral decomposition of $\nabla^2 E(\Q^*)$, 
\begin{equation}
    \nabla^2 E(\Q^*)=\sum_{i=1}^{i=m} \lambda_i \_v_i^* {\_v_i^*}^\top, |\lambda_1|\leqslant |\lambda_2|\leqslant\cdots \leqslant |\lambda_m|.
\end{equation}
Note that $\nabla E(\Q^*)=0$, consequently,
\begin{equation}
    \J(\Q^*,{\_v}^*_{j_1},\cdots,{\_v}^*_{j_k})=\frac{\partial(\dot{\Q}^*,\dot{\_v}^*_{j_1},\cdots,\dot{\_v}^*_{j_k})}{\partial(\Q^*,\_v_{j_1}^*,\cdots,\_v_{j_k}^*)}=\begin{pmatrix}
        \J_{\Q^*} & \mathbf{0} &\mathbf{0} & \cdots & \mathbf{0}\\
        \J_{1\Q^*} & * &  \mathbf{0}& \cdots & \mathbf{0}\\
        \J_{2\Q^*} & * &  *& \cdots & \mathbf{0}\\
        \vdots &\vdots & \vdots& &\vdots \\
        \J_{k\Q^*}&* &* &\cdots &*
    \end{pmatrix}
\end{equation}
\begin{equation}
    \J_{\Q^*}=-\left (\I-\sum_{i=j_1}^{i=j_k}2\_v^*_i {\_v_i^*}^\top\right)\sum_{i=1}^{i=m} \lambda_i \_v_i^* {\_v_i^*}^\top=\sum_{i=1}^{i=m} (-1)^{\alpha(i)} \lambda_i \_v_i^* {\_v_i^*}^\top,
\end{equation}
where $j_i,i=1,\cdots,k$ are the subscripts corresponding to negative eigenvalues and $\alpha$ is the indicator function of the set $\{j_i,i=1,\cdots,k\}$ (If there exists an element in the set $\{j_1,\cdots,j_k\}$ which is equal to $i$, $\alpha(i) = 1$, otherwise, $\alpha(i) = 0$). Thus, Cond$_2(\J(\Q^*)) \geqslant$ Cond$_2(\J_{\Q^*})=|\lambda_m|/|\lambda_1|$, which is relatively large due to the smallness of $\lambda_1$, i.e., the $k$-saddle dynamics is stiff when the iteration point is close to $\Q^*$. Consequently, the saddle dynamics offers high impedance to $\_v_1^*$. In fact, the saddle dynamics \eqref{eq: SD} is a special gradient method, and it exhibits the ``jagged phenomenon", i.e., the iteration point will slowly move along the eigenvector corresponding to the smallest absolute eigenvalue. The convergence rate is largely dependent on the degree of separation between $|\lambda_1|$ and $|\lambda_m|$. These numerical difficulties motivate us to develop a suitable numerical method to accelerate \eqref{eq: SD}.

The large stiffness of \eqref{eq: SD} necessitates a stable scheme. The linear term in \eqref{eq: SD} is implicitly discretized  for numerical stability. The nonlinear term, $|\Q|^2\Q$, is also semi-implicitly discretized in time direction as $|\Q^n|^2\Q^{n+1}$ for better numerical stability. The term $|\Q^n|^2\Q^{n+1}$ is very beneficial for solving linear equations in the semi-implicit scheme, because it is a positive definite term of the diagonal elements. Instead of re-generating unstable eigendirections with the gradient type dynamics in \eqref{eq: SD}, we apply a single-step Locally Optimal Block Preconditioned Conjugate Gradient (LOBPCG) method \cite{Knyazev2001TowardTO} to calculate the unstable eigendirections, and  the Hessians are also approximated by dimers \cite{2019High}. The semi-implicit scheme is given by,

\begin{equation}
    \left\{
    \begin{aligned}
		\frac{\Q_{n+1}-\Q_n}{\Delta t_n}=& \Delta_{\delta x} \Q_{n+1}-\lambda^2\left(\frac{A}{2C}\Q_{n+1}+\frac{1}{2}|\Q_{n}|^2\Q_{n+1}-\frac{B}{2C}\left({\Q_{n}}^2-\frac{|\Q_{n}|^2}{3}\I \right)\right) \\
		&+(2\sum_{i=1}^k {\_v}_{n,i}{\_v}_{n,i}^\top)\nabla_{\delta x} E(\Q_n),\\
		\text{Renew } \_v_{n,i} & \text{ as } \_v_{n+1,i} \text{ with single-step LOBPCG} , \  i=1,2,\cdots,k.\\
    \end{aligned}
    \right.
	\label{eq: Runge}
\end{equation}
This semi-implicit scheme is stable enough and it allows us to choose a large step size, which suffices for our purpose. We use finite difference methods to estimate the spatial derivatives in \eqref{eq: Runge} with mesh size $\delta x=1/32$. We have tested that the solutions are not sensitive to smaller choices of $\delta x$ by refining the mesh size.

The convergence rate is still slow due to the small eigenvalue, even with a large time step. We use Newton's method to accelerate the tail convergence, i.e., when the gradient is large, the saddle dynamics is used to ensure that $\Q_n$ falls into the basin of attraction of $\Q^*$, and then Newton's method pushes $\Q_n$ to $\Q^*$, with a higher convergence rate. However, Newton's method requires solving a large sparse ill-conditioned linear system, $R_n=\nabla^2 E(\Q_n) \delta \Q +\nabla E(\Q_n)=0$, at each step, 
and we hence, choose the Inexact-Newton method, i.e., give $R_n$ a tolerance $\Vert R_n \Vert\leqslant \eta_n \Vert \nabla E(\Q_n) \Vert$ with $\eta_n<1$. When the linear system is not very ill-conditioned (Cond$_2$($\nabla^2 E(\Q_n)\leqslant 10^{8}$), it can be solved within this tolerance by iteration methods, e.g., the generalized minimal residual method (GMRES) and symmetric successive over-relaxation method (SSOR). A small $\eta_n$ achieves faster convergence but leads to more expensive computational costs to solve the linear system. When $\Q_n$ is not too close to $\Q^*$, the matrix is not heavily ill-conditioned and we can solve the linear system more exactly to accelerate the convergence and keep $\Q_n$ in the basin of attraction of $\Q^*$; when $\Q_n$ is close to $\Q^*$, we choose a larger $\eta_n$ to save computational cost. Combining these considerations, we choose $\eta_n=\min(C, \bar{\eta}_n),\bar{\eta}_n=\frac{
1}{1+100\Vert \nabla E(\Q_n) \Vert}$ , and $0<C<1$ is a constant to guarantee at least linear convergence rate. In our numerical calculations, the calculation speed is sensitive to the choice of $C$, a small $C$ is more efficient when $h$ is relatively small. 

\begin{figure}[hbtp]
    \centering
    \includegraphics[width=.99\textwidth]{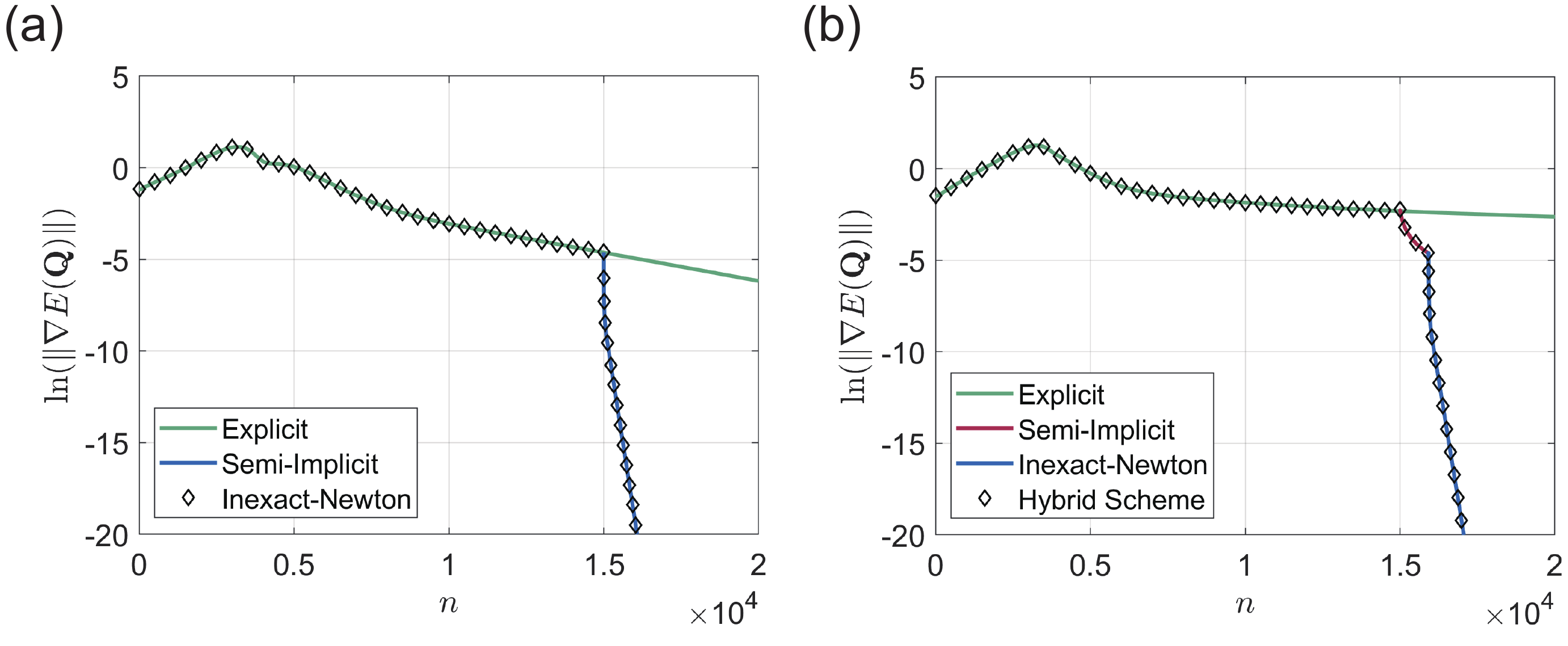}
    \caption{The upward search for finding an index-1 sample 3D LdG critical point D-B-D from the stable D state by following the 1-SD with (a) $h=1$ and (b) $h=2$. The parameters in \eqref{eq: Hybrid} are $\mu=10^{-2},N=15000,C=0.99$.}
    \label{fig:3}
\end{figure}

We solve five large sparse linear systems (five degrees of freedom) in \eqref{eq: Runge} at every time step, and the single-step LOBPCG needs another $4k$ derivative evaluations, which is computationally expensive, particularly for finding higher-index saddle points. 
Fortunately, we can use the explicit scheme combined with the Barzilai-Borwein step size \cite{BB} to save the computational cost at the beginning of the iteration. Thus, we use the explicit system, combined with the semi-implicit scheme and the Inexact-Newton method to propose the final hybrid numerical scheme:
\begin{equation}
\begin{cases}
    \text{The explicit scheme of \eqref{eq: SD}}, \  \Vert \nabla E(\Q_n)\Vert \geqslant \mu \text{ and } n\leqslant N,\\

    \text{The semi-implicit scheme \eqref{eq: Runge}},\ \Vert \nabla E(\Q_n)\Vert \geqslant \mu \text{ and } n>N,\\

    \Q_{n+1} =\Q_{n} +\delta \Q, \Vert \nabla^2 E(\Q_n) \delta \Q +\nabla E(\Q_n) \Vert \leqslant \eta_n\Vert \nabla E(\Q_n)\Vert \ ,\text{Otherwise},\\
\end{cases}
\label{eq: Hybrid}
\end{equation}
where $N$ is a step parameter to automatically identify the stiffness of \eqref{eq: SD}, and $\mu$ is chosen to be small enough to ensure the convergence of the Inexact-Newton method. For the well-conditioned case (Fig. \ref{fig:3}(a)), the iteration point can reach $\Vert \nabla E(\Q_n)\Vert<\mu$ within the step parameter, by means of the explicit scheme alone, and then Inexact-Newton method pushes convergence to the saddle point. For ill-conditioned cases (Fig. \ref{fig:3}(b)), the explicit scheme cannot achieve $\Vert \nabla E(\Q_n)\Vert<\mu$ within the step parameter, and the semi-implicit scheme is used to achieve $\Vert \nabla E(\Q_n)\Vert<\mu$ followed by the Inexact-Newton method 
to complete tail convergence. For example, when the cuboid height $h=1$, the CPU time is larger than 10000 seconds if we only use explicit scheme, while the CPU time for hybrid scheme is 2240 seconds (Table \ref{table:1}). For $h=2$, the CPU time is larger than 60000 seconds if we only use explicit or semi-implicit scheme, while the CPU time for hybrid scheme is 13678 seconds.

\begin{table}[hbtp]
    \small
    \renewcommand\arraystretch{1.2}
    \centering
    \caption{The CPU time (second) for each component of the hybrid scheme in Fig. \ref{fig:3}.}
    \label{table:1}
	\begin{tabular}{c|c c c}  
		\hline  
          & Explicit & Semi-implicit& Inexact-Newton  \\
        \hline
          $h=1$ & 1341  & 0 & 899   \\

         $h=2$ & 2552  & 5211 & 5915  \\
        \hline
    \end{tabular}
    \vspace{12pt}
\end{table}

\subsection{Construction of the solution landscape}
The solution landscape is an umbrella term used to describe the collection of unstable saddle points and stable critical points of the LdG free energy. Crucially, the solution landscape contains penetrating information about the pathways between critical points: how high-index saddle points are connected to low-index saddle points, and eventually to index-0 stable critical points, noting that not all critical points can be connected.

Following the discrete SD dynamics \eqref{eq: Hybrid}, we can construct the solution landscape without tuning initial guesses, by two algorithms: the downward search that enables us to search for connected index-$s$ saddle points from known index-$k$ ($k>s$) saddles $(\Q,\_v_1,\_v_2,\cdots,\_v_k)$, two typical choices for the initial guess of $s$-SD are $(\Q \pm \epsilon \_v_{s+1},\_v_1,\_v_2,\cdots,\_v_s)$; the upward search to find connected index-k saddle points from known index-$s$ ($k>s$) saddles $(\Q,\_v_1,\_v_2,\cdots,\_v_s,\bar{\_v}_{s+1},\cdots,\bar{\_v}_k)$, where $\bar{\_v}_i,i=s+1,\cdots,k$ are stable eigenvectors of $\Q$ and two typical choices for the initial guess of the $k$-SD are $(\Q \pm \epsilon \bar{\_v}_{s+1},\_v_1,\_v_2,\cdots,\_v_s,\bar{\_v}_{s+1},\cdots,\bar{\_v}_k)$ \cite{yin2020construction,yin2021searching}. In the next section, we present our numerical results, based on this hybrid numerical scheme for solution landscapes.

\section{Results} \label{results}
\subsection{$z$-invariant LdG critical point (A-A-A)}
\begin{figure}[hbtp]
    \centering
    \includegraphics[width=.99\textwidth]{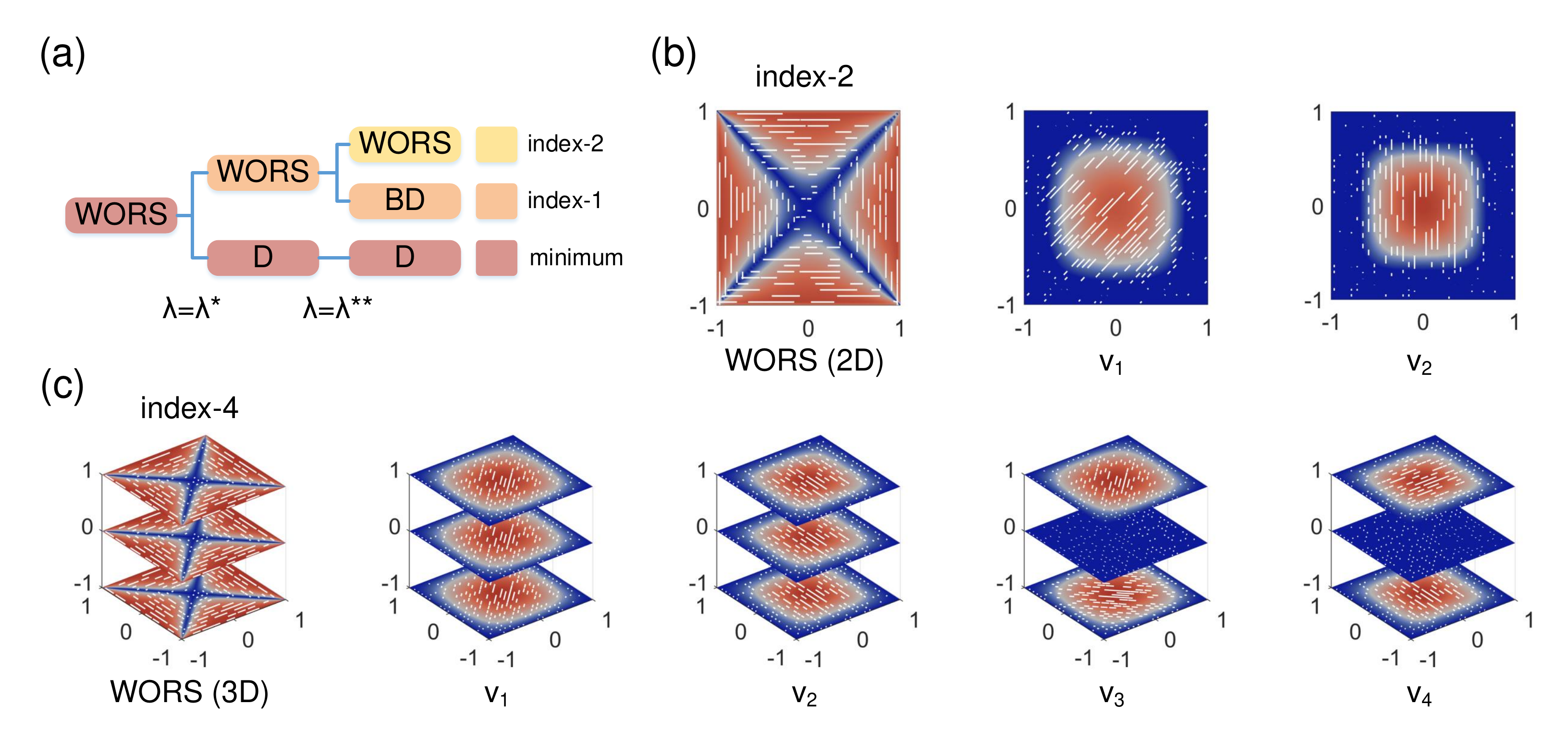}
    \caption{(a) The bifurcation diagram of 2D WORS as a function of $\lambda$. (b) The 2D WORS computed on $\Omega$ (the color bar is $\lambda_3-B/6C$), and its unstable eigenvectors at $\lambda^2=15$. (c) The 3D WORS with its unstable eigenvectors $v_1,\cdots,v_4$ with $\lambda^2=15,h=1$.}
    \label{fig:4}
\end{figure}

In this section, we show that there are differences between the 2D and 3D cases, even when restricted to $z$-invariant LdG critical points. A 3D $z$-invariant LdG critical point can be defined by $\Qvec (x, y, z) = \Qvec_{2D} (x, y)$ for $(x,y)\in \Omega; -h \leqslant z \leqslant h$, where $\Qvec_{2D}$ is a 2D LdG critical point on the bottom slice, $z=-h$. 
We take the Well Order Reconstruction Solution (WORS) as an example to illustrate the relation between the index of 3D $z$-invariant LdG critial point and the associated 2D LdG critial point. In \cite{kralj2014order}, the authors study LdG critical points on a square domain, with edge length $\lambda$ and tangent boundary conditions (consistent with \eqref{Dirichlet bc} on the lateral surfaces). For $\lambda$ small enough, the WORS is the unique LdG critical point, and hence the unique energy minimizer (see Fig. \ref{fig:4}). One can numerically show that the smallest eigenvalue of the Hessian of the LdG energy, at the WORS critical point, on $\Omega$, is strictly decreasing with increasing $\lambda$. 
As shown in Fig. \ref{fig:4}(a), at the first bifurcation point, $\lambda = \lambda^*$, the 2D WORS becomes an index-$1$ saddle point and bifurcates into two stable D solutions along the two unstable directions $+\_v_1$ and $-\_v_1$ in Fig. \ref{fig:4}(b). For the D solutions, the nematic director is almost aligned along one of the square diagonals. At the second bifurcation point $\lambda = \lambda^{**}$, the index-1 WORS becomes index-2 with two unstable eigenvectors, $\_v_1$ and $\_v_2$, and bifurcates into two boundary distortion (BD) solutions along the two unstable directions $+\_v_2$ and $-\_v_2$ in Fig. \ref{fig:4}(b).

\begin{table}[hbtp]
    \small
    \renewcommand\arraystretch{1.2}
    \centering
    \caption{The index of 2D and 3D ($h=1$) WORS versus $\lambda^2$. $N_{z-variant}$ is the number of unstable $z$-variant eigenvector of WORS, i.e., the index of 3D WORS minus the index of 2D WORS.}
    \label{table:2}
	\begin{tabular}{c|c c c c c c c c }  
		\hline  
         $\lambda^2$ & 2 & 7& 10 & 12 & 15& 19  & 22 & 30 \\
        \hline
          index of 2D WORS & 0 & 1 & 2 & 2 & 2 & 4 & 4 & 4 \\
          index of 3D WORS & 0 & 1& 2 & 3 & 4 & 6 & 7 & 10 \\

         $N_{z-variant}$ & 0 & 0& 0 & 1 & 2 & 2 & 3 & 6 \\
        \hline
    \end{tabular}
    \vspace{12pt}
\end{table}
Next, we consider the 3D WORS as an example of a $z$-invariant 3D LdG critical point (see Fig. \ref{fig:4}(c)) for all $\lambda$ and $h$, with the boundary conditions specified in (\ref{Dirichlet bc}). The 3D WORS is the global energy minimizer for sufficiently small $\lambda$ \cite{canevari_majumdar_wang_harris}. By analogy with the 2D case, the 3D WORS loses stability as $\lambda$ increases, for a fixed $h$. In fact, for a fixed $\lambda$ and $h=1$, we numerically observe that the index of the 3D $z$-invariant WORS is always greater than or equal to the index of the 2D WORS, because the eigenvectors of a 2D LdG critical point are also the eigenvectors of the corresponding $z$-invariant 3D LdG critical point. For example, the 3D WORS is an index-4 saddle point at $\lambda^2=15,h=1$ with two $z$-invariant unstable eigenvectors, $v_1$ and $v_2$, as in the 2D case. However, the 3D WORS can also accommodate unstable $z$-variant eigenvectors like $\_v_3$ and $\_v_4$, see Fig. \ref{fig:4}(c). As $\lambda$ increases, the 3D $z$-invariant WORS critical point has an increasing number of $z$-variant eigenvectors (which cannot be accommodated in the 2D case) and hence, this intuitively explains why the 3D $z$-invariant WORS has a higher Morse index than its 2D counterpart, for $\lambda$ is large enough, some of which is tabulated in Table \ref{table:2}.

\subsection{$z$-variant LdG critical points constructed by 2D pathways (A1-B-A2)}\label{sec:the solution landscape}

\begin{figure}[hbtp]
    \centering
    \includegraphics[width=.8\textwidth]{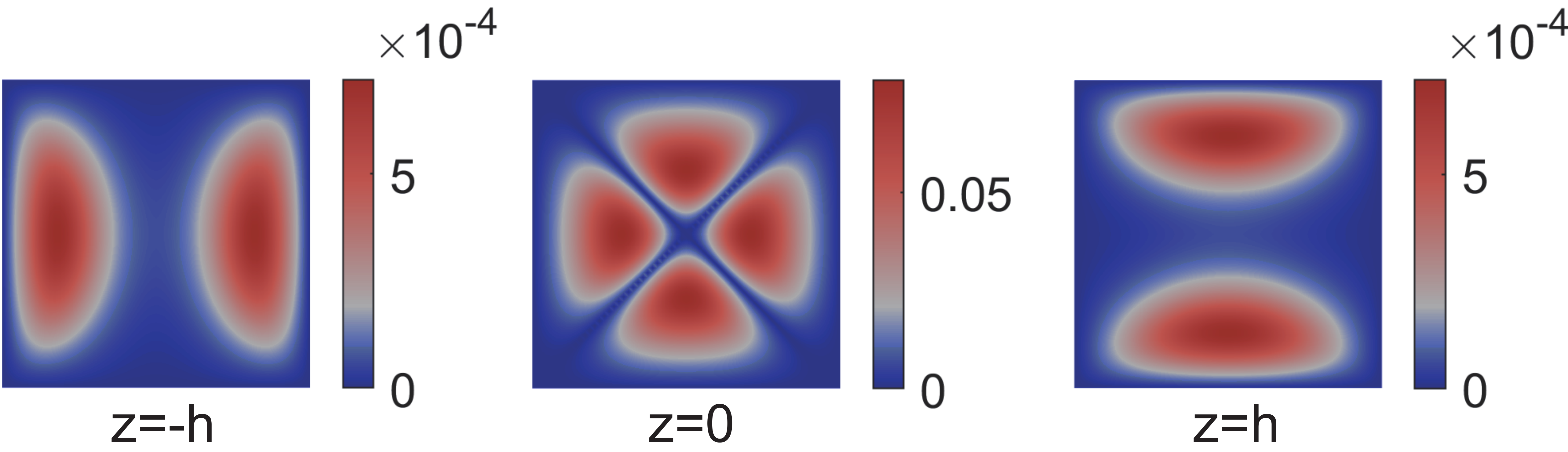}
    \caption{From left to  right are the plots of $\text{Tr}((\Q_1-\Q_2)^2)/2$, where $\Q_1$ are the slice plots on $z = -h,0,h$ of the 3D LdG critical point B-W-B, respectively, at $\lambda^2=30$ and $h=3$, and $\Q_2$ are the corresponding 2D critical points  (BD$_1$, WORS, and BD$_2$) at $\lambda^2=30$.}
    \label{fig:new}
\end{figure}

This section is devoted to constructing 3D $z$-variant critical point, A1-B-A2, by pathways on the corresponding 2D solution landscape, A1 $\rightarrow$ B $\rightarrow$ A2. 
The 2D slices located at $z = -h$, $0$, $h$ of a 3D $z$-variant solution, A1-B-A2, are not true critical points in 2D, but they are good approximations to the corresponding 2D critical points (see Fig. \ref{fig:new} where we plot the differences between the slice profiles and the corresponding 2D LdG critical points on $z=-h, 0, h$). We use this nomenclature for convenience, to study the relationship between 2D pathways on square domains and 3D $z$-variant LdG critical points on a 3D cuboid.

\begin{figure}[hbtp]
    \centering
    \includegraphics[width=.99\textwidth]{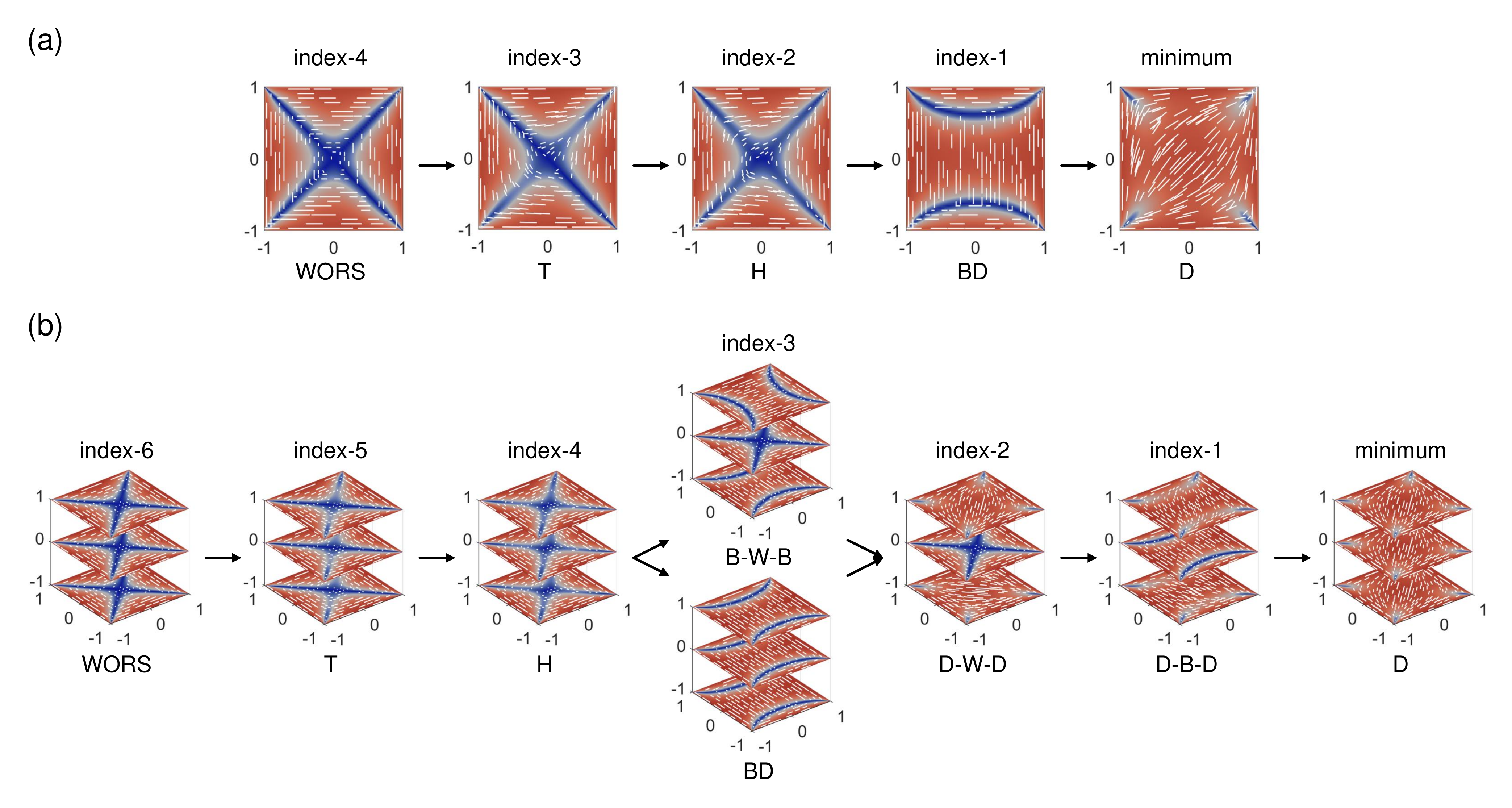}
    \caption{The solution landscape on (a) 2D square and (b) 3D cube at $\lambda^2=19$. The arrow from the higher-index to the lower-index saddle implies that the lower-index 3D solution can be found from the higher-index solution by following a downward search. 
    B and W are the shorthand of BD and WORS, respectively.}
    \label{fig:5}
\end{figure}

In Fig. \ref{fig:5}, we plot some numerical examples to this effect. Here, the 2D solution landscape can be summarised as WORS $\rightarrow$ T $\rightarrow$ H $\rightarrow$ BD $\rightarrow$ D (Fig. \ref{fig:5}(a)), where WORS is the parent saddle point with the highest Morse index and D is an index-$0$ stable diagonal solution. 
Considering the corresponding solution landscape on a cuboid or a 3D well with $\lambda^2=19$ and $h=1$ (Fig. \ref{fig:5}(b)), we have 3D $z$-invariant LdG critical points: WORS, T, H, BD and D. The Morse indices of the $z$-invariant critical points are ordered consistently with their 2D counterparts.
 The $z$-invariant 3D saddle points are connected to $z$-variant 3D LdG critical points and in some cases, we can use the pathways on the 2D solution landscape to heuristically explain the emergence and connectivity of the $z$-variant 3D LdG critical points. For example, the pathway between the two $z$-variant 3D LdG critical points, B-W-B to D-W-D  can be explained in terms of the pathway between the unstable BD to the stable D on the top and bottom slices, and D-W-D to D-B-D pathway can be explained in terms of the pathway from the unstable WORS to the lower-index BD solution on the middle slice, on the corresponding 2D solution landscape.
\begin{figure}[hbtp]
    \centering
    \includegraphics[width=.99\textwidth]{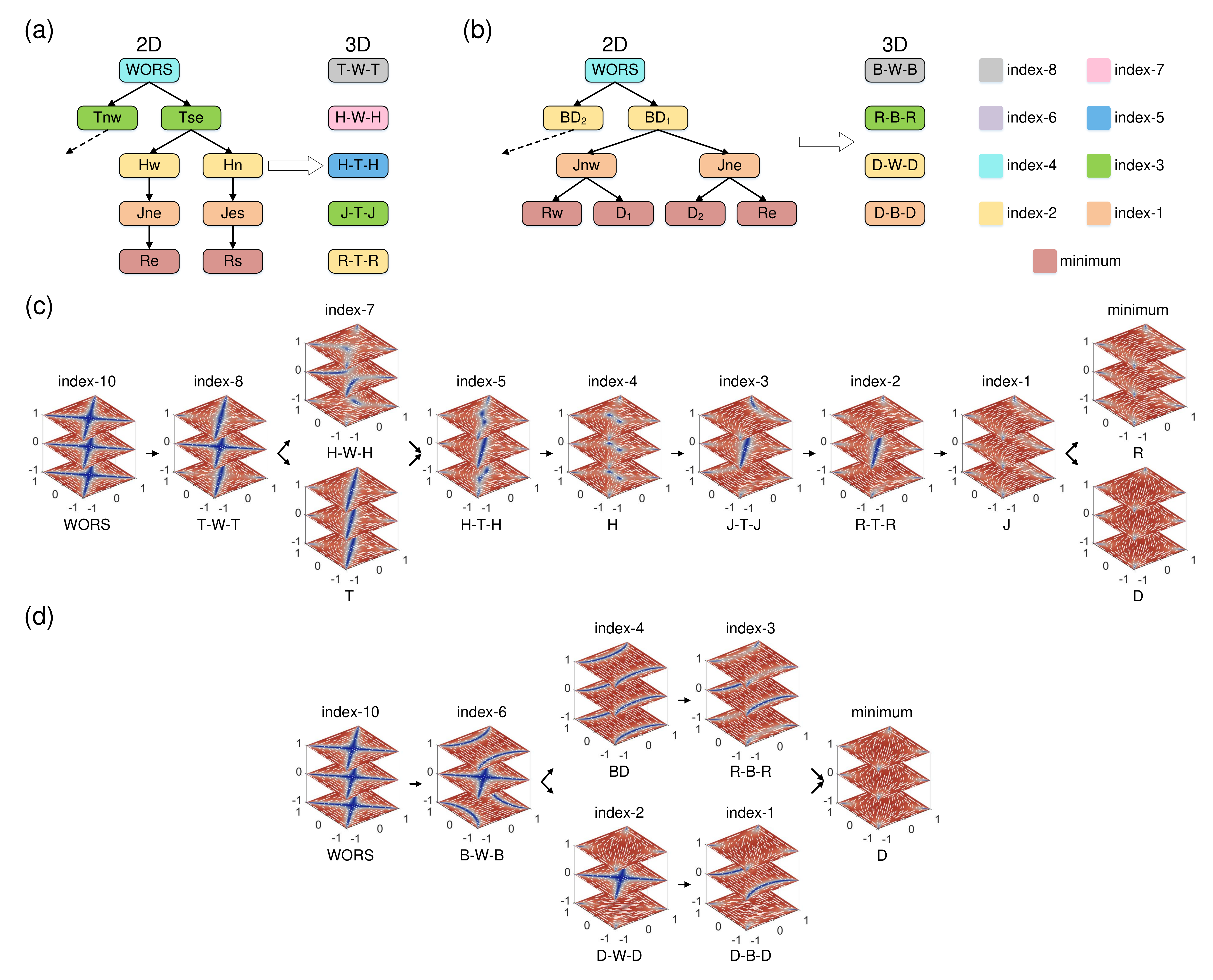}
    \caption{(a-b) are two pathway maps in the 2D solution landscape and the 3D solutions found by them at $\lambda^2=30$. (c-d) are the corresponding 3D configurations constructed by (a-b). The colors of the node specify the Morse indices of saddle points. The subscripts (e=east, w=west, n=north, s=south) distinct rotationally equivalent solutions. The symmetrical part in the pathway maps is omitted by the dashed arrows.}
    \label{fig:6}
\end{figure}

These numerical results suggest that the $z$-variant LdG critical points usually accommodate two lower-index 2D solutions on the top and bottom surfaces accompanied by a higher-index 2D saddle point on $z=0$ and the crucial question is - can we use pathways between distinct 2D LdG critical points as a database to construct  3D $z$-variant LdG critical points? The answer is affirmative and we use two 2D pathway maps, WORS $\rightarrow$ T $\rightarrow$ H $\rightarrow$ J $\rightarrow$ R and WORS $\rightarrow$ BD $\rightarrow$ J $\rightarrow$ D (R), to construct two branches of 3D solutions at $\lambda^2=30$ in Fig. \ref{fig:6}. However, we also observe $z$-variant 3D LdG critical points which cannot be mapped to pathways on the 2D solution landscape.

As shown in Fig. \ref{fig:6}(a), at $\lambda^2=30$, we have four 2D T states which are index-3, and the pathway between them passes through the index-4 WORS. We stack this pathway along the $z$-axis as an initial condition for our numerical algorithm and obtain a 3D $z$-variant LdG critical point, T-WORS-T, by using the SD. Similarly, we can obtain 3D $z$-variant LdG saddle points, H-W-H, H-T-H, J-T-J, and R-T-R from the following pathways on the 2D solution landscape:  
$\text{H}_\text{n}\rightarrow \text{WORS} \rightarrow \text{H}_\text{s}$, $\text{H}_\text{n}\rightarrow \text{T}_\text{se} \rightarrow \text{H}_\text{w}$, $\text{J}_\text{es}\rightarrow \text{T}_\text{se} \rightarrow \text{J}_\text{ne}$, and $\text{R}_\text{s}\rightarrow \text{T}_\text{se} \rightarrow \text{R}_\text{e}$, respectively. Combined with the $z$-invariant solutions, we show the relatively complete 3D solution landscape in Fig. \ref{fig:6}(c).
As we progress from the parent state (WORS) of the 3D solution landscape, the indices of the 2D LdG critical points on the top and bottom typically decrease or the index of the 2D middle slice decreases. For example, the $z$-invariant WORS is an index-10 saddle point and relaxes to an index-8 T-W-T by relaxing the top and bottom surfaces to the 2D T profile. The T-W-T relaxes to an index-7 H-W-H critical point, by relaxing the T states to the H states, or relaxing the middle slice to T results in a 3D $z$-invariant index-7 T state. The H-W-H relaxes the middle slice to T, or the $z$-invariant T state relaxes the top and bottom slices to the H state, so that both of these 3D LdG critical points relax to an index-5 H-T-H. The H-T-H state has two line defects running throughout the cuboid, that smoothly
interpolate between the $+1/2$ and $-1/2$ planar point defects on the top and bottom surfaces respectively, and this cannot be observed in 2D. Similarly, 
the 2D solution landscape in Fig. \ref{fig:6}(b) is used to construct the 3D solution landscape in Fig. \ref{fig:6}(d).

However, not all pathways on 2D solution landscapes lead to 3D LdG critical points, e.g. we cannot construct the $z$-variant J-B-J state from the 2D pathway $\text{J}_\text{nw}\rightarrow\text{BD}_\text{1}\rightarrow\text{J}_\text{ne}$ at $\lambda^2=30,h=1$, whereas we are able to find it for larger $\lambda$ or $h$. The Euler--Lagrange equation on the rescaled domain $[-1,1]^3$ is
\[\frac{1}{\lambda^2}\partial^2_{x} \Q +\frac{1}{\lambda^2}\partial^2_{y} \Q + \frac{1}{h^2\lambda^2}\partial^2_{z} \Q =\left(\frac{A}{2C}\Q-\frac{B}{2C}\left(\Q^2-\frac{tr(\Q^2)}{3}\I \right)+\frac{1}{2}tr(\Q^2)\Q\right).\]
As the height $2\lambda h$ of the cuboid increases, the effect of the term $\frac{1}{h^2\lambda^2}\partial^2_{z} \Q$
is weakened, and the system can better accommodate $z$-variant solutions. This raises the fundamentally interesting question of whether we can provide algorithmic recipes for using pathways on 2D solution landscapes for systematically constructing 3D LdG critical points, in the $\lambda^2 \to \infty$ or $h \to \infty$ limit.

\subsection{Multiple--layer solutions (A1-B-A2-B-A1)}

\begin{figure}[hbtp]
    \centering
    \includegraphics[width=.99\textwidth]{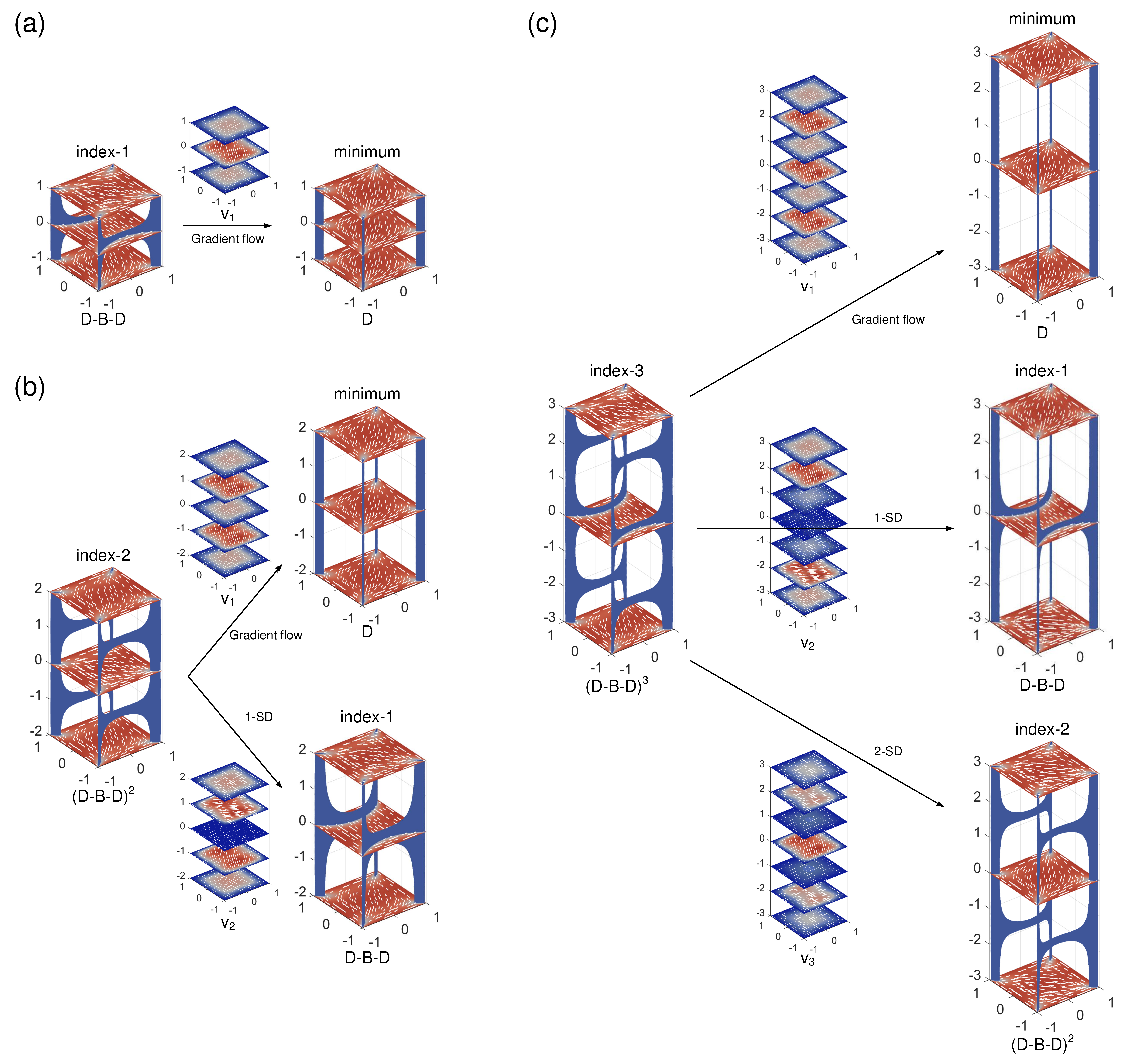}
    \caption{The D-B-D type solutions and the downward search along their unstable eigenvectors with (a) one layer (b) two layers and (c) three layers with $\lambda^2=30$.}
    \label{fig:8}
\end{figure}
In this section, we study the relationship between the Morse indices of multiple-layer solutions and the number of layers. We can construct multiple-layer solutions (A-B-A)$^n$ by stacking  $n$ blocks of $z$-variant 3D LdG critical points, A-B-A, on top of each other. We use the D$_2$-B-D$_1$ saddle point (labelled as D-B-D), to illustrate this point in Fig. \ref{fig:8}. Recall that there are two diagonal, D$_1$ and D$_2$, solutions, since there are two square diagonals. Based on the numerical observations in Fig. \ref{fig:8}, where the configurations and connections between multiple-layer solutions (D-B-D)$^i$, $i = 1,2,3$ are shown, we have the following conjectures. The multiple-layer solution, (D-B-D)$^n$ is an index-$n$ saddle point with unstable eigen-directions $\_v_1,\cdots,\_v_n$. If $n$ is even, then we have the same diagonal state (D$_1$ or D$_2$) at the top and bottom; if $n$ is odd, we necessarily have different diagonal states on the top and bottom. With the disturbance of $\_v_i,1\leqslant i \leqslant n$, the (D-B-D)$^n$ relaxes to (D-B-D)$^{(i-1)}$ saddle point (assuming the $z$-invariant D solution to be (D-B-D)$^0$), by following the $(i-1)$-SD.  The energy of (D-B-D)$^{(i-1)}$ is lower than that of (D-B-D)$^n$, for all $1\leqslant i\leqslant n$. It is an open question as to whether these numerical observations can be proven or generalized to other multiple-layer saddle points.

\subsection{The transition pathways}

\begin{figure}[hbtp]
    \centering
    \includegraphics[width=.65\textwidth]{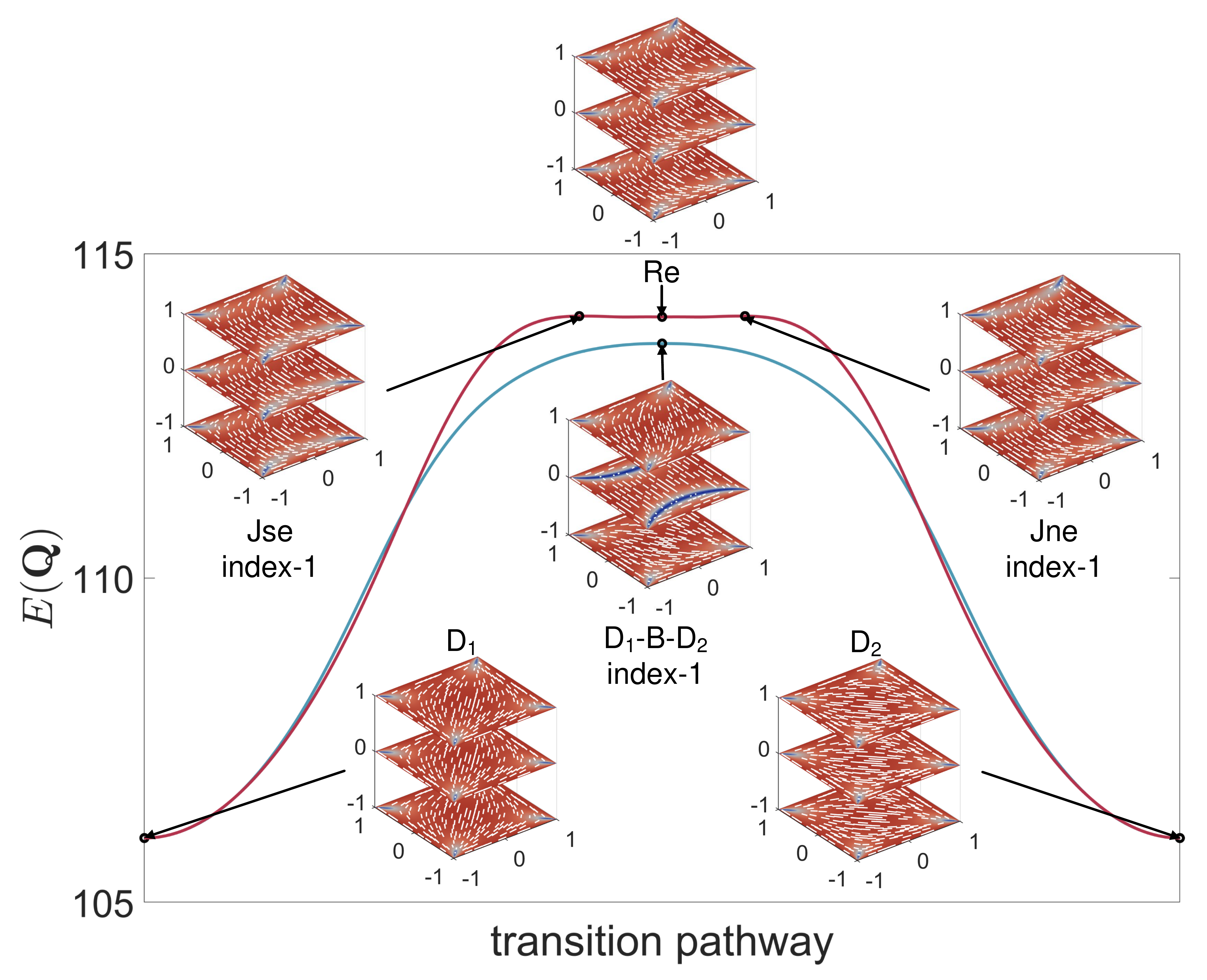}
    \caption{Transition pathways between two dual D states, D$_1$ and D$_2$, with $\lambda^2=30$, $h=1$. The vertical axis is the LdG energy \eqref{energy_b}, and the horizontal axis describes the transition pathway.
}
    \label{fig:7}
\end{figure}
We find two transition pathways between the $z$-invariant stable D$_1$ and D$_2$ states, via $z$-invariant and $z$-variant transition states, respectively.
The transition state is the index-1 saddle point and plays a key role in determining the energy barrier of transition pathways \cite{zhang2007morphology,zhang2021construction,yin2021transition}. In the 2D case, D$_1$ and D$_2$ correspond to diagonal states with the director along one of the two square diagonals. The two diagonal states cannot be connected by a single transition state in 2D for large domain size \cite{yin2020construction,kusumaatmaja2015free}. The switching between the two diagonal states must go through a two-stage transition that involves a metastable rotated state and two distinct transition states
, i.e. the pathway sequence D$_1$ $\rightarrow$ J$_\text{se}$ $\rightarrow$ R$_\text{e}$  $\rightarrow$ J$_\text{ne}$ $\rightarrow$ D$_2$. In fact, this transition pathway also exists in 3D (Fig. \ref{fig:7}). 
In 3D, we also find another switching mechanism between the diagonal states by passing through a $z$-variant 3D LdG saddle point, index-1 D-B-D, for which the system breaks the 2D restriction. We believe the second pathway is more likely to occur in practice than the first one for the appropriate domain dimensions, since it has a lower energy barrier and avoids the risk of being trapped into a metastable  state. However, the second pathway only exists with a relatively large cuboid height. Thus, for small cuboid height, the transition is achieved by rotating liquid crystal molecules/directors in the $xy$-plane i.e. via a $z$-invariant transition pathway. For larger values of $h$, the system prefers to switch between D$_1$ to D$_2$ by utilizing
the third dimension and the transition state is the $z$-variant D-B-D critical point. This is an example of how we can control switching mechanisms in bistable systems \cite{inbook}
by manipulating the experimental setup.

From our numerical results, we speculate that if we generalize our work to 3D wells with a hexagonal or arbitrary polygonal cross-sections, we will have $z$-variant 3D LdG critical points which could act as transition states for new transition pathways between globally stable $z$-invariant LdG critical points, so that this example on a cuboidal domain is generic in nature.

\subsection{The effects of $\lambda^2$ and $h$ on the 3D solution landscape} \label{section:the bifurcation diagram as a function of lambda square and h}
\begin{figure}[hbtp]
    \centering
    \includegraphics[width=.95\textwidth]{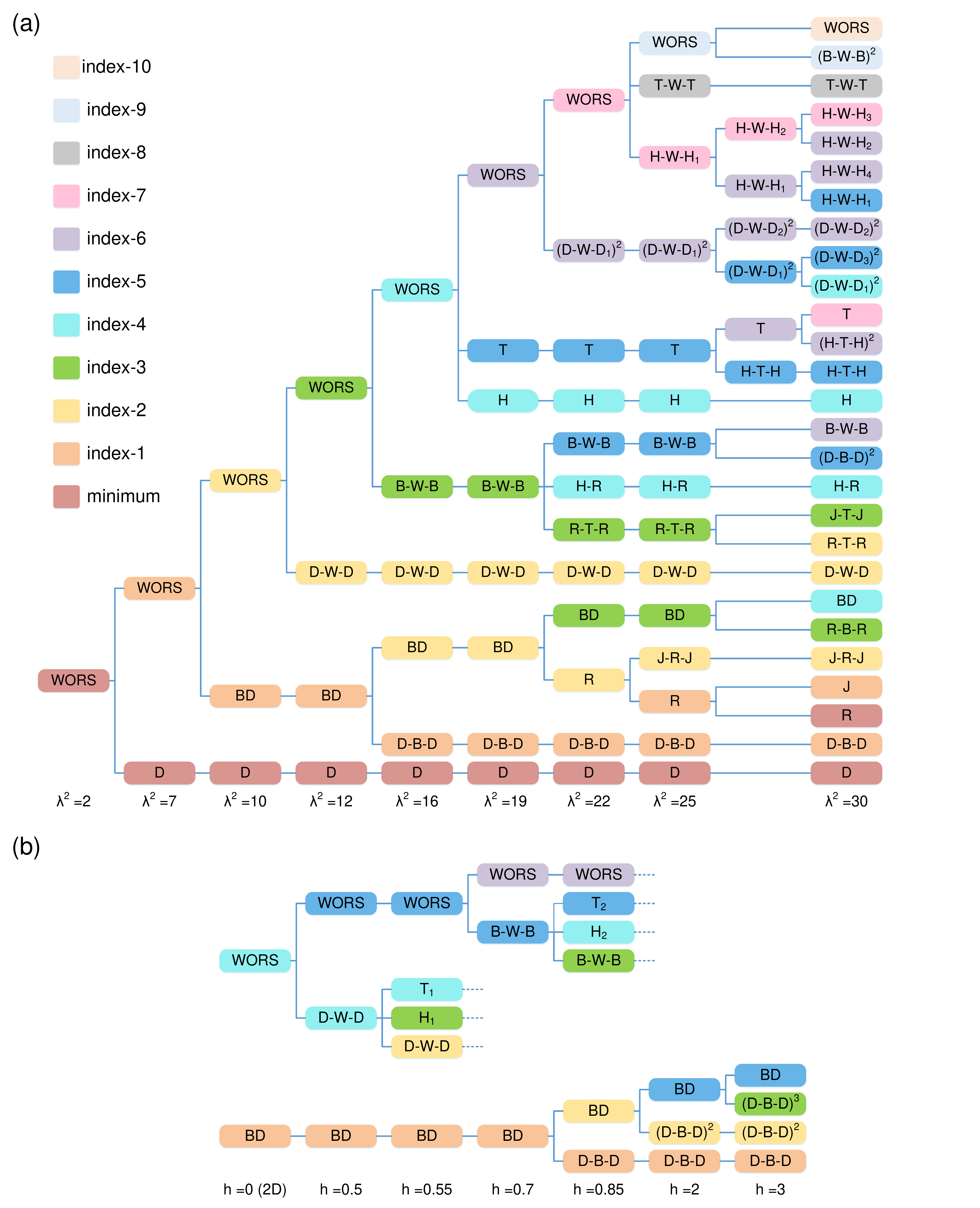}
    \caption{(a) Bifurcation diagram as a function of $\lambda^2$ at $h=1$. (b) Bifurcation diagram as a function of $h$ at $\lambda^2=19$. Each small rectangle represents a solution as shown in Fig. \ref{fig:5} and Fig. \ref{fig:6} and the color represents the Morse index. The subscript of solutions, e.g. H-W-H$_1$ and H-W-H$_2$, is used to distinguish solutions with similar defect configurations and the superscript denotes a multiple-layer solution, e.g., $(\text{D-B-D})^2$ has two layers of D-B-D. Each T-junction represents a pitchfork bifurcation. We omit some subsequent  bifurcations by using the dashed line for brevity.}
    \label{fig:9}
\end{figure}

In this section, we make some preliminary observations about the effects of $\lambda^2$ and $h$ on the 3D solution landscape. 
We make these observations more precise by computing part of the bifurcation diagram, for solutions of (\ref{EL_b}) subject to (\ref{Dirichlet bc}), as a function of $\lambda^2$ with $h=1$, in Fig. \ref{fig:9}(a).  We track the Morse indices of the solutions in Sec. \ref{sec:the solution landscape}, since a change in the Morse index signals the onset of a bifurcation \cite{shi2021nematic}. As $\lambda$ increases, the domain is able to accommodate more defects; the indices of $z$-invariant solution increase; it is easier to find $z$-variant states constructed by 2D transition pathways or multiple-layer solutions; and the solution landscape is more complicated. For example, at $\lambda^2=12$, an index-2 WORS bifurcates into an index-2 $z$-variant D-W-D which cannot be observed in the 2D case. At $\lambda^2 = 22$, the index-6 WORS bifurcates into an index-6 2-layer solution: (D-W-D)$^2$.

We also track part of the bifurcation diagram as a function of $h$ in Fig. \ref{fig:9}(b), to study the effect of the cuboid height on the solution landscape, at $\lambda^2=19$. We only focus on the WORS and BD branches for simplicity.  
As $h$ increases, on the one hand, the index of $z$-invariant solutions, like BD and WORS increases. On the other hand, the energetic penalty of distortions in the $z$-direction decreases which informally explains why we observe more $z$-variant LdG saddle points, like D-W-D and B-W-B, with decreasing Morse indices, i.e., enhanced stability. This is corroborated by Fig. \ref{fig:2}, where the smallest eigenvalue of the Hessian of the LdG energy at D-B-D increases to zero as $h\to \infty$, which indicates that the $z$-variant 3D solution is more stable with increasing $h$.

\subsection{Escaped solutions}
\begin{figure}[hbtp]
    \centering
    \includegraphics[width=.97\textwidth]{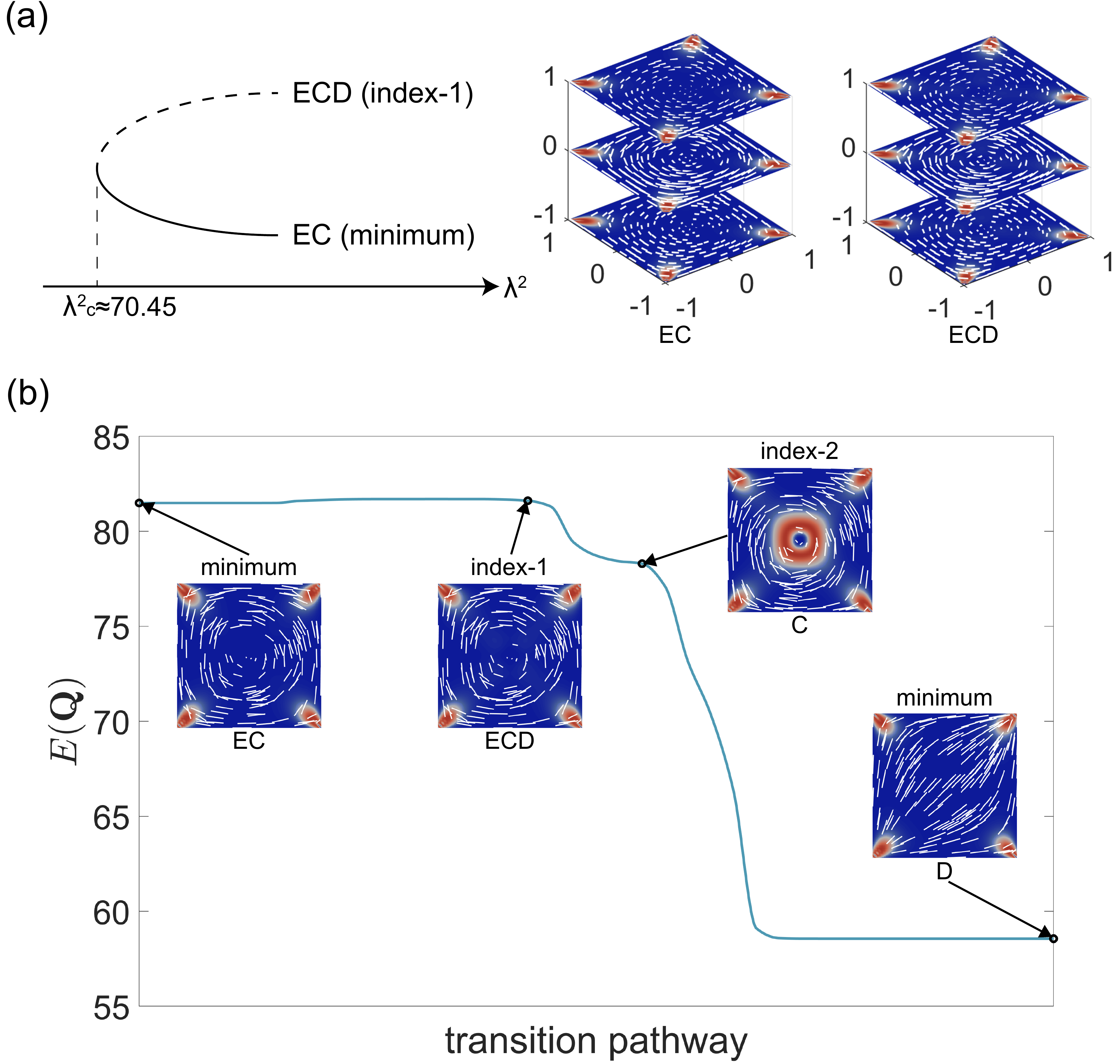}
    \caption{(a) The saddle-node bifurcation between EC and ECD, and the cross-section of the EC and ECD at $\lambda^2$=74. (b) The transition pathway beween EC and D in 2D case at $\lambda^2$=74. Color bar is the biaxiality parameter $\beta^2=1-6\text{tr}(\Q^3)^2/(\text{tr}(\Q^2))^3$ (see Sec. 2). The vertical axis is the 2D LdG energy and the horizontal axis describes the transition pathway.}
    \label{fig:10}
\end{figure}
Recall the five degrees of freedom of the LdG critical points denoted by $q_1, \ldots, q_5$ in \eqref{eq:5-degree}.
For the numerical results presented in the previous sections, we have $q_4=q_5=0$ and $q_3$ is largely a constant, i.e. this physically means that $\Q$ has a fixed eigenvector in the $\zhat$ direction and the remaining two eigenvectors are in the $xy$-plane. 
This raises the interesting question - do we have LdG critical points, with the choice of boundary conditions in (\ref{Dirichlet bc}) and natural boundary conditions on $z=\pm h$, that exploit the full five degrees of freedom?
In \cite{wang2019order}, the authors demonstrate two escaped solutions with non-zero $q_4$ and $q_5$, and non-constant $q_3$, on a 2D square domain with an isotropic concentric square inclusion \cite{wang2019order}. We build on the work in \cite{wang2019order} and add Gaussian perturbation to the $z$-invariant C state (see Fig. \ref{fig:10}(b) or \cite{yin2020construction}) to construct a suitable initial condition that converges to two escaped solutions in our framework. These escaped solutions exist for relatively large $\lambda^2$, and they are $z$-invariant stable states, labelled as escaped +1 center (EC+) and escaped -1 center (EC-), where $\pm1$ indicates that the director rotates by $\pm \pi$ radians anticlockwise around the center. They have non-zero $q_4$ and $q_5$ profiles and are energetically degenerate, and hence, we only study the EC state with $+1$ center (Fig. \ref{fig:10}(a)). Using the upward search, we can find an index-1 ECD from the stable index-0 EC state, for $\lambda^2>70.45$. In fact, the stable EC and index-1 ECD emerge from a saddle-node bifurcation at $\lambda^2 \approx 70.45$, without bifurcation connections with the WORS branch. As $\lambda^2$ increases, the EC state is always stable whilst the Morse index of the ECD increases, and bifurcates into multiple $z$-invariant and $z$-variant escaped solutions. We do not analyze this further in this paper, largely because the structure of this escaped branch is similar to the WORS branch in Sec. \ref{results}.

Since the EC and ECD LdG critical points are $z$-invariant, their cross-sections exist as critical points in 2D cases, for the same value of $\lambda^2$. The 2D ECD critical point is an index-1 saddle point, while the 2D EC is a metastable  state since it has higher energy than the D state at $\lambda^2=74$. We investigate the transition pathway between EC and D in the 2D case at $\lambda^2=74$ (Fig. \ref{fig:10}(b)). The transition state is the index-1 ECD state, and the energy barrier ($E(\text{ECD})-E(\text{EC})$) is low, so that the system can easily escape from the trap of the metastable EC state. 
It is noteworthy that the transition pathway passes through an index-2 C state, which is connected to the WORS and the C state has only three degrees of freedom. In other words, in order to transition from the EC state (which  exploits five degrees of freedom) to the D state (which exploits three degrees of freedom, or two degrees of freedom if $q_3$ is constant as in \eqref{Q_c}), the escaped directors are pulled back into the $xy$-plane, and the transition pathway 
goes from escaped branch to the WORS branch, and finally, reaches the D state.

\section{Discussion and conclusion}
We study critical points of a LdG free energy on a 3D cuboid with Dirichlet tangent boundary conditions on lateral surfaces and natural boundary condition on top and bottom surfaces, in terms of two geometry-dependent variables: the  cuboid size $\lambda$, and the height $h$. 
First, we design a hybrid numerical scheme to discretize and accelerate the saddle dynamics. 
Our notable findings include (i) $z$-variant LdG critical points that depend on the third dimension, (ii) new pathways between energy minimizers mediated by $z$-variant critical points which are inaccessible in 2D, (iii) multiple-layer LdG critical points and (iv) novel stable escaped solution branches. Essentially, the solution landscapes become increasingly complicated as $\lambda^2$ and $h$ increases. We find intimate connections between pathways on 2D solution landscapes (for 2D domains in a reduced LdG framework) and $z$-variant 3D LdG critical points. Whilst our work is not exhaustive, we can typically construct $z$-variant 3D LdG critical points by interpolating between two distinct 2D reduced LdG critical points, and the interpolation usually involves a third higher-index 2D critical point on the middle slice of the cuboid. Of course, not all pairs of 2D reduced LdG critical points are compatible; we typically need 2D dual critical points that are connected by a pathway on the 2D solution landscape, to construct the $z$-variant 3D counterpart. As $\lambda \to \infty$, we speculate that we could use the entire database of dual 2D critical points to construct $z$-variant 3D LdG critical points. Hence, reduced 2D studies have value in higher dimensions too.

There are numerous open questions stemming from this work. 
For example, can we have 3D LdG critical points that interpolate between an escaped solution and a non-escaped solution? Are there other disconnected LdG critical points on a 3D cuboid and if so, how to find them? 
Our working domain is a cuboid with a square cross-section, but these methods could be easily generalized to a 3D well with an arbitrary 2D cross-section e.g. rectangle, hexagons etc. In fact, on a rectangle, we lose the degeneracy between different critical points e.g. the dual BD states are not energetically degenerate on a rectangle and the 2D WORS branch divides into two unconnected branches \cite{shi2021nematic}. Thus, some of the solutions in this paper, e.g., BD-WORS-BD will have a different structure for a 3D well with a rectangular cross-section. Finally, we could work with weak tangential anchoring on the lateral well surfaces, as opposed to Dirichlet conditions. In particular, the nematic director profile on the lateral surfaces is constrained to be one-dimensional by \eqref{Dirichlet bc}, which severely constrains the solution space. Weak boundary conditions allow for more freedom on the lateral surfaces, which naturally adds further possibilities for the corresponding solution landscapes. Finally, there is scope for rigorous asymptotic analysis in the $\lambda^2\rightarrow\infty$ limit \cite{han2020reduced}, and we expect close correspondence with some of the analytic results in \cite{majumdar2006} in this limit.

\section*{Acknowledgements}
This work was supported by the National Key R\&D Program of China 2021YFF1200500,
the National Natural Science Foundation of China 12225102 and 12050002, and the Royal Society
Newton Advanced Fellowship awarded to L. Zhang and A. Majumdar. A. Majumdar is supported by a Leverhulme Research Project Grant RPG-2021-401, a Leverhulme International Academic Fellowship IAF-2019-009. Y. Han gratefully
acknowledges the support from a Royal Society Newton International Fellowship and the Leverhulme Research Project Grant RPG-2021-401. J.~Yin is supported by the National Research Foundation, Singapore (project No.~NRF-NRFF13-2021-0005).
\section*{References}
\bibliographystyle{unsrt}
\bibliography{references}
\end{document}